\documentclass[twocolumn,prx,10pt,showpacs]{revtex4-1}
\usepackage[utf8]{inputenc}
\usepackage{graphicx}
\usepackage{amsmath}
\usepackage{amssymb}
\usepackage{bm}
\usepackage{color}
\usepackage{xcolor}
\usepackage{braket}

\usepackage{hyperref}
\hypersetup{colorlinks=true,linkcolor=blue,anchorcolor=blue,citecolor=blue,filecolor=blue,urlcolor=blue,bookmarksnumbered=true,pdfview=FitB}

\graphicspath{{./Figures/}}

\begin{document}

\title{Zero-energy Andreev bound states from quantum dots in proximitized Rashba nanowires}

\author{Christopher Reeg, Olesia Dmytruk, Denis Chevallier, Daniel Loss, and Jelena Klinovaja}
\affiliation{Department of Physics, University of Basel, Klingelbergstrasse 82, CH-4056 Basel, Switzerland}

\date{\today}
\begin{abstract}
We study an analytical model of a Rashba nanowire that is partially covered by and coupled to a thin superconducting layer, where the uncovered region of the nanowire forms a quantum dot. We find that, even if there is no topological superconducting phase possible, there is a trivial Andreev bound state that becomes pinned exponentially close to zero energy as a function of magnetic field strength when the length of the quantum dot is tuned with respect to its spin-orbit length such that a resonance condition of Fabry-Perot type is satisfied. In this case, we find that the Andreev bound state remains pinned near zero energy for Zeeman energies that exceed the characteristic spacing between Andreev bound state levels but that are smaller than the spin-orbit energy of the quantum dot. Importantly, as the pinning of the Andreev bound state depends only on properties of the quantum dot, we conclude that this behavior is unrelated to topological superconductivity. To support our analytical model, we also perform a numerical simulation of a hybrid system while explicitly incorporating a thin superconducting layer, showing that all qualitative features of our analytical model are also present in the numerical results.
\end{abstract}

\maketitle

\section{Introduction} 
Majorana bound states (MBSs) are zero-energy quasiparticles that emerge at the boundaries of one-dimensional (1D) topological superconductors \cite{Kitaev:2001,Alicea:2012,Beenakker:2013}, with their non-Abelian statistics and topological protection making these states highly sought for potential applications in topological quantum computing \cite{Kitaev:2001}. The theoretical proposal to engineer a topological superconductor hosting MBSs in hybrid semiconductor-superconductor structures \cite{Oreg:2010,Lutchyn:2010} has subsequently received extensive experimental attention. Following several experiments that observed zero-bias peaks in the differential conductance of such a hybrid system \cite{Mourik:2012,Deng:2012,Das:2012,Churchill:2013,Finck:2013}, much emphasis was placed on improving the quality of the semiconductor-superconductor interface \cite{Lutchyn:2018}. This led to the development of epitaxial interfaces between thin shells of aluminum (Al) and both hexagonal nanowires (either InAs \cite{Chang:2015} or InSb \cite{Gazibegovic:2017}) and two-dimensional electron gases (InAs) \cite{Kjaergaard:2016,Shabani:2016}, providing a very strong proximity effect into the semiconductor. Experiments in both 1D \cite{Deng:2016,Zhang:2018,Vaitiekenas:2018,Deng:2018,deMoor:2018} and 2D \cite{Suominen:2017,Nichele:2017} setups consisting of a proximitized nanowire coupled to a lead via a normal region that is left uncovered by the superconductor (which we henceforth refer to as a ``quantum dot") have demonstrated an ability to reliably generate coalescing Andreev bound states (ABSs) that give rise to zero-bias conductance peaks persisting over large ranges of applied magnetic field strength, consistent with what one might expect to observe in the presence of MBSs.

With the advent of hybrid experimental systems that can be tuned \cite{Vaitiekenas:2018,deMoor:2018} to the strong-proximity regime, there has been a recent theoretical emphasis placed on realistic treatments of the proximity effect in this limit while accounting for the small thickness of the superconducting shell. Studies based on an analytical tunneling Hamiltonian approach \cite{Reeg:2017_3,Reeg:2018,Reeg:2018_2} have shown that all material parameters (such as effective mass, $g$-factor, and spin-orbit strength) of the semiconductor get significantly renormalized toward their corresponding values in the superconductor due to the strong proximity coupling (similarly to studies assuming the superconductor to be infinitely large \cite{Sau:2010prox,Stanescu:2010,Potter:2011,Tkachov:2013,Zyuzin:2013,Cole:2015,vanHeck:2016,Hell:2017,Stanescu:2017,Reeg:2017_2}) and that all subbands within the semiconductor experience very large shifts in their effective chemical potentials that are highly dependent on the geometry of the superconducting shell. The combination of these two effects makes it difficult to realize a topological phase in the strong-coupling limit before destroying superconductivity in the shell. Additionally, several subsequent studies based on a numerical self-consistent Schr\"{o}dinger-Poisson approach \cite{Antipov:2018,Woods:2018,Mikkelsen:2018} have also shown that a great degree of fine tuning of experimental parameters is required to realize a topological phase in the presence of a strong coupling to a thin superconducting shell. Because it has been demonstrated that a hard superconducting gap, which is required for the topological protection of qubits encoded in MBSs, can only be experimentally induced in the nanowire in the strong-coupling limit \cite{deMoor:2018}, it is possible that the observed zero-bias conductance peaks do not originate from MBSs.

Recently, an alternative explanation that the observed zero-bias peaks are due to topologically trivial ABSs originating within the quantum dot has been put forth \cite{Liu:2017,Ptok:2017}. These trivial ABSs have subsequently been shown to also reproduce many experimental signatures of topological MBSs beyond zero-bias conductance peaks, such as quantized conductance, $4\pi$-periodic Josephson effect, and a high degree of nonlocality as measured by the coupling to a normal lead \cite{Setiawan:2017,Moore:2018,Vuik:2018,Avila:2018,Hansen:2018,Elsa2}. However, previous studies have been restricted to numerical simulations and thus relatively little is known about precisely which properties of the system are responsible for stabilizing low-energy ABSs, though a common suggestion is that smooth variations in either the chemical potential or pairing potential are required \cite{Setiawan:2017,Moore:2018,Vuik:2018,Avila:2018} (note that such smooth variations can also stabilize low-energy ABSs in the absence of a normal dot region \cite{Kells:2012,Fleckenstein:2018,Aseev:2018}). Additionally, the ABSs have always been shown to evolve into topological MBSs \cite{Pascal1, Pascal2, Cayao:2015} as the strength of the magnetic field is increased, indicating that they could simply be a precursor to a true topological phase. Finally, all previous studies have considered effective (strict) 1D models that fail to account for the significant modifications of the nanowire by the  proximity effect and the strong dependence of the system properties on the geometry of the superconducting layer.

\begin{figure}[t!]
\includegraphics[width=\linewidth]{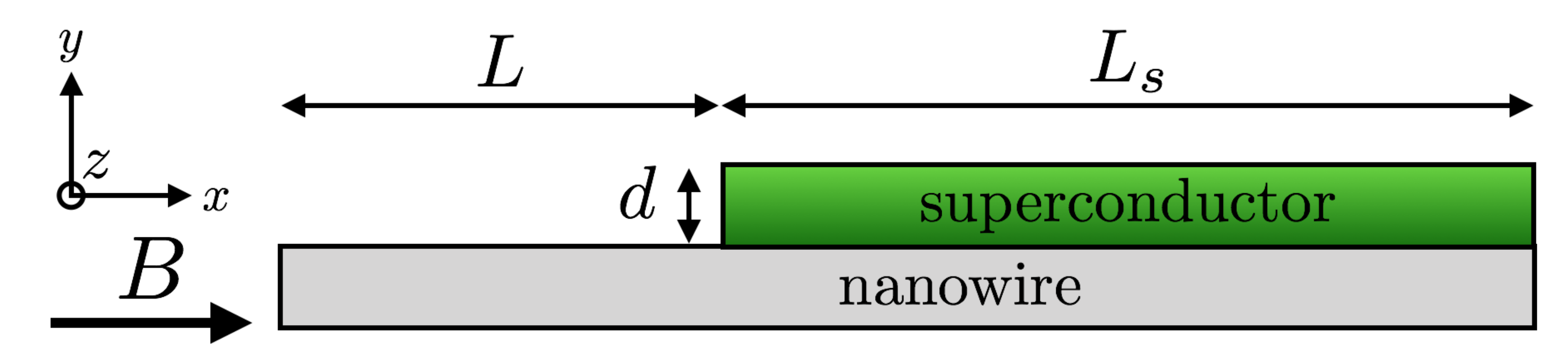}
\caption{A semiconducting Rashba nanowire (gray) of length $L+L_s$ is partially covered by a thin $s$-wave superconductor (green) of thickness $d$ and length $L_s$. Part of the nanowire is left uncovered by the superconductor, forming a normal quantum dot region of length $L$. A uniform magnetic field $B$ is applied parallel to the nanowire axis.
}
\label{fig:Fig1}
\end{figure}

In this paper, we first consider a minimal analytical model of a quantum dot/nanowire junction (see Fig.~\ref{fig:Fig1}) in which we take the proximitized region of the nanowire to have no spin-orbit interaction (SOI). We show that even in such a model in which there is never a topological phase (due to the fact that there is no SOI in the proximitized section of the wire), it is still possible to have a low-energy ABS that persists over a large range of magnetic field strength. This low-energy ABS is thus purely a property of the quantum dot and not related to topological superconductivity.  Provided that the spin-orbit energy $E_{so}$ is the largest energy scale within the dot region, we find that a low-energy ABS arises if the chemical potential is tuned within the Zeeman gap and if a Fabry-Perot-like resonance condition relating the spin-orbit length and the dot length ($L$) is satisfied. In particular, we find that the presence of such an ABS is insensitive to all other device properties and does not rely on smooth variations of any parameter \cite{Hansen:2018}. We also show that the relevant energy scale determining the Zeeman energy ($\Delta_Z$) needed to pin the ABS exponentially close to zero energy is the characteristic spacing between ABS levels $\alpha/L$ (where $\alpha/\hbar$ is the Fermi velocity in the dot region), such that the ABS remains pinned near zero energy for $\alpha/L\lesssim\Delta_Z\lesssim E_{so}$. As the low-energy ABS is a property only of the quantum dot, our analytical simplification of taking the proximitized section to have no SOI is not crucial. Therefore, the physical mechanism that we discuss giving rise to the ABS is present in much more general cases than the experimentally relevant strong-coupling limit.

We additionally perform numerical simulations of a nanowire/superconductor hybrid system in which a region of the nanowire is left uncovered by the superconductor. As we explicitly incorporate a superconducting layer with finite thickness, our numerical model accounts for both the strong proximity coupling and the strong dependence of the proximitized region on the geometry of the superconductor, two crucial aspects of the system that have been neglected in previous works. Consistent with our analytical model, we find a low-energy ABS by tuning the spin-orbit length to a resonant value with respect to the dot length $L$, with no topological phase transition occurring due to the strong proximity coupling. Crucially, we find that the presence of this ABS is insensitive to the thickness of the superconducting layer and the details of the proximity effect; as the properties of the proximitized region of the nanowire, and thus the topological phase transition, are highly dependent on these two quantities \cite{Reeg:2017_3,Reeg:2018,Reeg:2018_2}, this result further demonstrates that the ABS is a property of the dot and unrelated to any properties of the superconducting region. Finally, we show that if the quantum dot is removed, there are no ABSs present in the system at energies far below the superconducting gap \cite{Huang:2018,Aseev:2018}.

The remainder of the paper is organized as follows. In Sec.~\ref{analytical}, we present and solve a minimal analytical model describing the nanowire/superconductor hybrid system. In Sec.~\ref{numerical}, we discuss a numerical simulation of the hybrid system while explicitly incorporating the superconducting layer. We show that all of the qualitative features of our analytical model are consistent with our numerical results. Our conclusions are given in Sec.~\ref{conclusion}.

\section{Analytical Calculation of Andreev Bound State Spectrum} \label{analytical}
We consider a system shown in Fig.~\ref{fig:Fig1}, where a Rashba nanowire is partially covered by a superconductor. A region of the nanowire of length $L$ is left uncovered and forms a normal quantum dot region with finite level spacing. To gain insight into the physical mechanism responsible for generating a low-energy ABS in such a system, we consider a minimal model in which the proximitized region of the nanowire has no SOI. While we focus on the case where a nanowire with intrinsic SOI is strongly coupled to a superconductor (thereby renormalizing the SOI via the proximity effect), we emphasize that such a model can apply to more general cases as well; e.g., the minimal model can also describe a nanowire with no intrinsic SOI that is weakly coupled to a superconductor, with local gates generating SOI only in the quantum dot region. Through an analytical solution of the ABS spectrum in this model, we find that a low-energy ABS is present if the chemical potential is tuned within the Zeeman gap and if the length of the nanowire region that is left uncovered by the superconductor (the quantum dot) is finely tuned with respect to the spin-orbit length to fulfill a resonance condition. Importantly, the presence of the ABS is sensitive only to properties of the dot and is therefore unrelated to topological superconductivity. 

It was found in Refs.~\cite{Reeg:2017_3,Reeg:2018} that a strong proximity coupling causes a significant renormalization of semiconducting material parameters and generally induces a large effective chemical potential shift in the nanowire, which can push the topological phase transition to prohibitively large magnetic field strengths. 
To capture these features of the proximity effect, we consider an effective one-subband model to describe the system shown in Fig.~\ref{fig:Fig1}, $H_\text{eff}=\int dx\,\Psi^\dagger(x)\mathcal{H}_\text{eff}\Psi(x)$, where the Hamiltonian density is given by ($\hbar=1$)
\begin{equation} \label{model}
\begin{aligned}
\mathcal{H}_\text{eff}&=\tau_3\biggl[-\partial_x\left(\frac{1}{2m(x)}\partial_x\right)-\mu(x)-\Delta_Z(x)\sigma_1\biggr] \\
	&+\frac{i}{2}[\alpha(x)\partial_x+\partial_x\alpha(x)]\sigma_3-\Delta(x)\tau_2\sigma_2,
\end{aligned}
\end{equation}
$\Psi(x)=[\psi_\uparrow(x),\psi_\downarrow(x),\psi^\dagger_\uparrow(x),\psi^\dagger_\downarrow(x)]^T$ is a Nambu spinor of operators $\psi_\sigma^\dagger(x)$ [$\psi_\sigma(x)$] that create (destroy) a particle of spin $\sigma$ at position $x$ in the nanowire, while $\sigma_{1,2,3}$ ($\tau_{1,2,3}$) are Pauli matrices acting in spin (particle-hole) space. Note that the derivative operator in Eq.~\eqref{model} acts on all quantities to its right, e.g., $\partial_x\alpha\Psi=(\partial_x\alpha)\Psi+\alpha(\partial_x\Psi)$, to ensure the Hermiticity of the Hamiltonian \cite{Reeg:2015,Klinovaja:2015}. We assume that all parameters take different values in the normal ($-L<x<0$) and superconducting ($x>0$) regions of the nanowire \cite{step}; note that we take the superconducting region to be semi-infinite for analytical calculations ($L_s\to\infty$). In particular, due to renormalization by the proximity effect, we neglect the Rashba SOI $\alpha$ and Zeeman energy $\Delta_Z=g\mu_BB/2$ (arising from an external magnetic field $B$ applied parallel to the nanowire axis, with $g$ the $g$-factor and $\mu_B$ the Bohr magneton) in the superconducting region, such that $\alpha(x)=\alpha\theta(-x)$ and $\Delta_Z(x)=\Delta_Z\theta(-x)$ [$\theta(x)$ is a Heaviside step function] \cite{neglect}. We neglect any orbital effects arising from the magnetic field \cite{olesia,serra,anton}. Pairing is induced only in the superconducting region of the nanowire, $\Delta(x)=\Delta\theta(x)$, while the mass is taken as $1/m(x)=(1/m)\theta(-x)+(1/m_s)\theta(x)$. The external magnetic field is incorporated in the superconducting region through the suppression of the bulk proximity gap, $\Delta=\Delta_0\sqrt{1-(B/B_c)^2}=\Delta_0\sqrt{1-(\Delta_Z/\Delta_Z^c)^2}$, where $B_c$ is the critical field at which superconductivity is destroyed (we additionally define the Zeeman energy induced in the quantum dot at the critical field as $\Delta_Z^c=g\mu_BB_c/2$). Finally, the chemical potential of the superconducting region is taken to be a large energy scale, $\mu(x)=\mu\theta(-x)+\mu_s\theta(x)$, where $\mu_s\gg\Delta$. We stress that our model [Eq.~\eqref{model}] does not have any topological phase regardless of the strength of the applied magnetic field due to the fact that $\alpha=0$ in the superconducting region.
\begin{widetext}
The normal quantum dot ($-L<x<0$) is described by the Bogoliubov-de Gennes (BdG) equation
\begin{equation} \label{BdGw}
\left[\left(-\frac{\partial_x^2}{2m}-\mu\right)\tau_3+i\alpha\partial_x\sigma_3-\Delta_Z\tau_3\sigma_1\right]\psi_n(x)=E\psi_n(x).
\end{equation}
To simplify the calculation, we assume that the spin-orbit energy $E_{so}=m\alpha^2/2$ is the largest energy scale in the normal region ($E_{so}\gg\Delta_Z,|\mu|$). A general solution to Eq.~\eqref{BdGw} to lowest order is then given by
\begin{equation} \label{solN}
\begin{aligned}
\psi_n(x)&=c_1\begin{pmatrix} v_{+} \\ -u_{+} \\ 0 \\ 0 \end{pmatrix}e^{-\Omega_{+}x/\alpha} +c_2\begin{pmatrix} u_{+} \\ -v_{+} \\ 0 \\ 0 \end{pmatrix}e^{\Omega_{+}x/\alpha}+c_3\begin{pmatrix} 1 \\ 0 \\ 0 \\ 0 \end{pmatrix}e^{i[2k_{so}+(E+\mu)/\alpha]x}+c_4\begin{pmatrix} 0 \\ 1 \\ 0 \\ 0 \end{pmatrix}e^{-i[2k_{so}+(E+\mu)/\alpha]x} \\
	&+c_5\begin{pmatrix} 0 \\ 0 \\ v_{-} \\ u_{-} \end{pmatrix}e^{-\Omega_{-}x/\alpha}+c_6\begin{pmatrix} 0 \\ 0 \\ u_{-} \\ v_{-} \end{pmatrix}e^{\Omega_{-}x/\alpha}+c_7\begin{pmatrix} 0 \\ 0 \\ 1 \\ 0 \end{pmatrix}e^{-i[2k_{so}-(E-\mu)/\alpha]x}+c_8\begin{pmatrix} 0 \\ 0 \\ 0 \\ 1 \end{pmatrix}e^{i[2k_{so}-(E-\mu)/\alpha]x},
\end{aligned}
\end{equation}
where $k_{so}=m\alpha$ is the spin-orbit momentum, $c_{1-8}$ are unknown coefficients to be found by imposing boundary conditions, $u_{\pm}=[\text{sgn}(E\pm\mu)/\sqrt{2}]\sqrt{1+ i\Omega_{\pm}/(E\pm\mu)}$, $v_{\pm}=(1/\sqrt{2})\sqrt{1- i\Omega_{\pm}/(E\pm\mu)}$, and $\Omega_{\pm}=\sqrt{\Delta_Z^2-(E\pm\mu)^2}$.

The superconducting region of the nanowire is described by the BdG equation
\begin{equation} \label{BdGs}
\left[\left(-\frac{\partial_x^2}{2m_s}-\mu_s\right)\tau_3-\Delta\tau_2\sigma_2\right]\psi_s(x)=E\psi_s(x).
\end{equation}
Focusing only on subgap energies $E<\Delta$, the general solution to Eq.~\eqref{BdGs} that decays in the limit $x\to\infty$ is given by
\begin{equation} \label{wfs}
\begin{aligned}
\psi_s(x)&=e^{-\Omega_\Delta x/v_F}\left[c_9\begin{pmatrix} u_\Delta \\ 0 \\ 0 \\ v_\Delta \end{pmatrix}e^{ik_Fx}+c_{10}\begin{pmatrix} 0 \\ u_\Delta \\ -v_\Delta \\ 0 \end{pmatrix}e^{ik_Fx}+c_{11}\begin{pmatrix} 0 \\ -v_\Delta \\ u_\Delta \\ 0 \end{pmatrix}e^{-ik_Fx}+c_{12}\begin{pmatrix} v_\Delta \\ 0 \\ 0 \\ u_\Delta \end{pmatrix}e^{-ik_Fx}\right],
\end{aligned}
\end{equation}
\end{widetext}
where $k_F=\sqrt{2m_s\mu_s}$ is the Fermi momentum, $v_F=\sqrt{2\mu_s/m_s}$ is the Fermi velocity, $u_\Delta(v_\Delta)=(1/\sqrt{2})\sqrt{1\pm i\Omega_\Delta/E}$, and $\Omega_\Delta=\sqrt{\Delta^2-E^2}$. Although we focus on a model in which topological superconductivity is not possible (as we have fixed $\alpha=\Delta_Z=0$ in the superconducting region), we note that Eq.~\eqref{wfs} is valid to leading order in the limit $\mu_s\gg E_{so,s},\Delta_{Z,s}$ even if $\alpha_s\neq0$ and $\Delta_{Z,s}\neq0$ in the superconducting region. Therefore, the following calculation applies to any general case for which $\mu_s\gg E_{so,s},\Delta_{Z,s}$. We show in Appendix~\ref{appA} that similar results to those presented in this section can also be found assuming the SOI and Zeeman energy to be spatially uniform, emphasizing that the parameter renormalization caused by the proximity effect does not play a crucial role.

To determine the spectrum of ABSs, we must impose three boundary conditions \cite{compose,Peter}. First, we impose that the wave function vanishes at the free end of the system, $\psi_n(-L)=0$. Second, the wave function should be continuous at $x=0$, $\psi_n(0)=\psi_s(0)$. And third, the quasiparticle current should be conserved, leading to the condition $\hat v_n\psi_n(0)=\hat v_s\psi_s(0)$, where $\hat v_n=(-i\partial_x/m)\tau_3-\alpha\sigma_3$ and $\hat v_s=(-i\partial_x/m_s)\tau_3$ are the velocity operators [note that this boundary condition can also be obtained by directly integrating Eq.~\eqref{model}]. These three boundary conditions can be arranged as a single equation of the form $M\vec c=0$, where $\vec c$ is a vector of unknown coefficients and $M$ is a matrix (which is too cumbersome to spell out here explicitly). This matrix equation has a nontrivial solution if
\begin{equation} \label{solvability}
\det(M)=0,
\end{equation}
and this solvability condition determines the spectrum of the system.

In the absence of a magnetic field ($\Delta_Z=0$), the ABS spectrum is given by solutions to the transcendental equation [Eq.~\eqref{solvability}]
\begin{equation} \label{ABSnoB}
\begin{aligned}
0&=(\alpha^2+v_F^2)\Omega_\Delta\cos\chi-2E\alpha v_F\sin\chi \\
	&-(v_F^2-\alpha^2)\Omega_\Delta\cos(2\bar k_{F}L),
\end{aligned}
\end{equation}
where $\chi=2EL/\alpha$ and $\bar k_F=(k_{F1}+k_{F2})/2$; $k_{F1}=2k_{so}+\mu/\alpha$ and $k_{F2}=\mu/\alpha$ are the two Fermi momenta of the Rashba spectrum. We note that Eq.~\eqref{ABSnoB} reproduces the known ABS spectrum $\Omega_\Delta\cos\chi=E\sin\chi$ in the limit of no Fermi velocity mismatch between the normal and superconducting regions ($\alpha=v_F$) \cite{deGennes:1963}. Fermi velocity mismatch acts as an effective sharp potential barrier at the interface ($x=0$) that introduces Fabry-Perot-like oscillations described by $\cos(2\bar k_FL)$. The number of ABSs lying within the gap is determined by the length of the quantum dot, as the spacing between ABS levels is given by the energy scale $\alpha/L$. Thus, for $\alpha/L\ll\Delta$, there are many subgap states, while for $\alpha/L\gg\Delta$, there is a single subgap state. To see that there is always at least one subgap ABS \cite{olesia2}, let us examine the limit $L/\alpha\to0$, where the ABS energy should be near the gap edge. Expanding $E=\Delta-\delta E$ (with $0<\delta E\ll\Delta$) and solving Eq.~\eqref{ABSnoB} gives
\begin{equation} \label{ABSnoBsol}
\sqrt{\frac{\delta E}{\Delta}}=\frac{\sqrt{2}v_F\alpha\chi_\Delta}{\alpha^2[1+\cos(2\bar k_{F}L)]+v_F^2[1-\cos(2\bar k_FL)]},
\end{equation}
where $\chi_\Delta=2\Delta L/\alpha$. As the right-hand side of Eq.~\eqref{ABSnoBsol} is positive definite (and $\ll1$), a solution to Eq.~\eqref{ABSnoB} always exists and there is always at least one spin-degenerate subgap ABS.

In the presence of a magnetic field, the spin-degenerate ABSs determined by Eq.~\eqref{ABSnoB} split,  and it is possible to have an ABS near zero energy even in the limit of a short dot ($\alpha/L\gg\Delta$, or equivalently $\chi_\Delta\ll1$). We now determine the conditions necessary for an ABS to remain pinned near zero energy over a large range of magnetic field strength. From the form of the wave function in Eq.~\eqref{solN}, we see that the Zeeman energy enters Eq.~\eqref{solvability} at low energies ($E\to0$) through the factors $\exp(\pm\sqrt{\Delta_Z^2-\mu^2}L/\alpha)$. Thus, it is clear that these factors can lead to an ABS energy that decays exponentially with $\Delta_Z$ provided that the chemical potential is tuned within the Zeeman gap (such that $|\mu|<\Delta_Z$). Additionally, the Zeeman energy at which the ABS becomes pinned near zero energy is minimized by tuning to $\mu=0$, in which case the ABS energy is pinned for $\Delta_Z\gg\alpha/L$. To find under what conditions it is possible to have a pinned ABS, let us expand Eq.~\eqref{solvability} to lowest order in the limits $E,|\mu|\ll\Delta,\Delta_Z,\alpha/L$ while keeping only the most relevant terms in the limit $\Delta_Z\gg\alpha/L$ \cite{inequality}. After expanding, we find a solvability condition given by
\begin{equation} \label{ABSB0}
\begin{aligned}
0&=4\Delta^2\left[2(v_F^2+\alpha^2)-(v_F^2-\alpha^2)\cos(2k_{so}L)e^{\chi_Z/2}\right]^2 \\
	&-E^2\bigl[16v_F^2\alpha^2+8v_F\alpha(v_F^2+\alpha^2)\chi_\Delta+(v_F^2-\alpha^2)^2\chi_\Delta^2\bigr]e^{\chi_Z} \\
	&+2\mu\Delta\chi_\Delta\sin(2k_{so}L)e^{\chi_Z/2}[2(v_F^4-\alpha^4) \\
	&-(v_F^2-\alpha^2)^2\cos(2k_{so}L)e^{\chi_Z/2}],
\end{aligned}
\end{equation}
where $\chi_Z=2\Delta_ZL/\alpha$. 

\begin{figure}[t!]
\includegraphics[width=\linewidth]{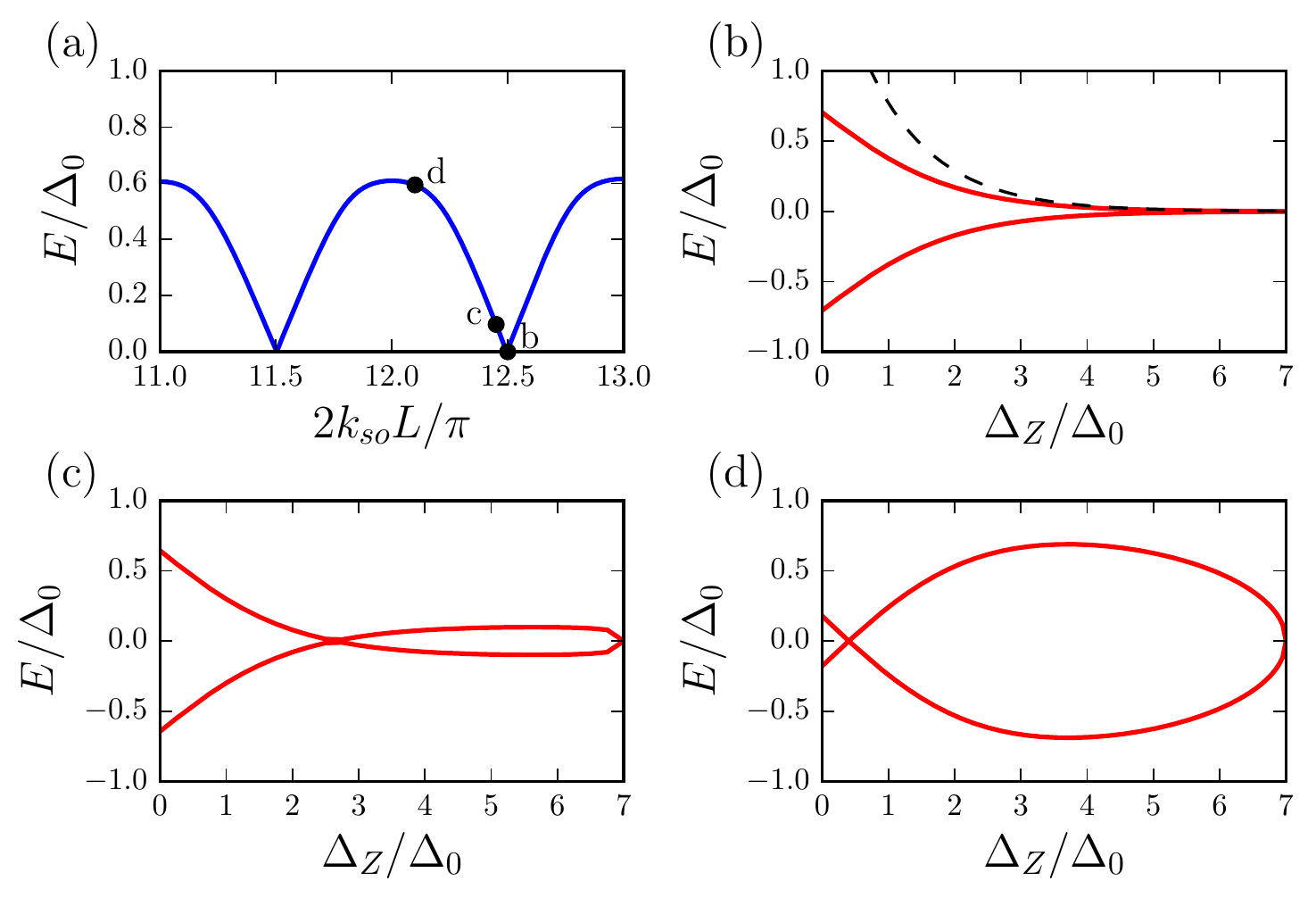}
\caption{\label{analytics} (a) The ABS energy,  found by numerically solving Eq.~\eqref{solvability} for $\Delta_Z=0.75\Delta_Z^c$, oscillates as $k_{so}L$ is varied. Points marked in the plot correspond to the values of $k_{so}L$ chosen in panels (b)--(d). (b) When tuning to the resonance condition $\cos(2k_{so}L)=0$, a subgap ABS energy is pinned exponentially close to zero as a function of Zeeman energy $\Delta_Z$. The dashed black curve corresponds to the asymptotic analytical expression given in Eq.~\eqref{EABS}. (c), (d) When $k_{so}L$ is tuned away from the resonance value, the ABS energy is no longer strictly pinned to zero. All plots were obtained for $\mu=0$, $v_F=10\alpha$, $L=604$ nm, $m=0.02m_e$, and $\Delta_Z^c/\Delta_0=7$.
}
\end{figure}

From Eq.~\eqref{ABSB0}, we see that $E=0$ is a valid solution for the ABS energy only for a single value of $\Delta_Z$, thus corresponding to a finely tuned zero-energy crossing. This behavior is very different than that of a semi-infinite topological superconducting nanowire, where a strict zero-energy MBS persists for all field strengths exceeding the topological phase transition.

While a strict $E=0$ solution is possible for only a single value of $\Delta_Z$, it is possible for a trivial ABS to persist near zero energy over a wide range of magnetic field strength. From Eq.~\eqref{ABSB0}, we see that this behavior is most pronounced in the presence of Fermi velocity mismatch  ($v_F\neq \alpha$) at $\mu=0$ when $\cos(2k_{so}L)=0$, corresponding to a finely tuned resonance condition determined solely by the spin-orbit length $1/k_{so}$ and the length of the quantum dot $L$. When the resonance condition is satisfied, the ABS energy is given by
\begin{equation} \label{EABS}
E=(2\alpha/L)e^{-\Delta_ZL/\alpha},
\end{equation}
where we have additionally assumed that $v_F\gg\alpha$ [and $(v_F/\alpha)\chi_\Delta\gg1$], though we note that this assumption changes only the prefactors in Eq.~\eqref{EABS} and does not affect the exponential behavior \cite{nomismatch}. 

If the system is slightly off-resonance with $\cos(2k_{so}L)\neq0$, a constant shift of the ABS energy is induced, given by $(\alpha/L)|\cos(2k_{so}L)|$, which remains small as long as $|\cos(2k_{so}L)|\ll\chi_\Delta$. If the system is on resonance $\cos(2k_{so}L)=0$ but is tuned away from $\mu=0$, this also induces a constant shift of the ABS energy given by $|\mu|/2$. While we consider in this section only small deviations away from $\mu=0$, we show numerically in Appendix~\ref{appA} that the pinning behavior can withstand larger deviations provided that one simultaneously alters the resonance condition. Additionally, although we obtain a sharp resonance condition in a calculation of the spectrum at zero temperature, we note that this resonance will be broadened by finite temperature and a coupling to external leads in a transport experiment; in this case, one should observe a zero-bias conductance peak if the broadening is larger than the energy of the ABS.

The results of our analytical calculation are summarized in Fig.~\ref{analytics}. In Fig.~\ref{analytics}(a), we show by solving Eq.~\eqref{solvability} numerically that the energy of the subgap ABS oscillates as a function of $k_{so}L$. Furthermore, we show in Fig.~\ref{analytics}(b) that if the system is tuned to the resonance condition $\cos(2k_{so}L)=0$ (assuming $\mu=0$), the ABS becomes pinned near zero energy as a function of Zeeman energy $\Delta_Z$ as predicted in Eq.~\eqref{EABS}. As $k_{so}L$ is tuned away from resonance, the ABS becomes less pinned as shown in Figs.~\ref{analytics}(c-d). Thus, within our analytical model we have shown that an ABS can be pinned near zero energy for $\alpha/L\ll\Delta_Z\ll E_{so}$. While these are rather stringent restrictions on the spin-orbit strength, we show in Appendix~\ref{appA} by numerically diagonalizing the model defined in Eq.~\eqref{model} that these inequalities can be loosened to $\alpha/L\lesssim\Delta_Z\lesssim E_{so}$ while retaining the same qualitative behavior obtained analytically.

This hierarchy of energy scales forms the central result of our paper, as it allows us to place bounds on the range of Zeeman fields over which a pinned ABS can be observed. It is also worth emphasizing that these bounds are consistent with experimental observations in epitaxial Al setups, where a zero-bias peak is typically observed for field strengths $\sim1$ T \cite{Deng:2016,Zhang:2018,Vaitiekenas:2018,Deng:2018}. Assuming a spin-orbit energy for a typical semiconducting (InAs or InSb) nanowire of about $E_{so}\sim1$ meV (which corresponds to a spin-orbit strength $\alpha\sim0.8$ eV \AA) and a $g$-factor $g\sim20$ \cite{Lutchyn:2018}, and given that a typical quantum dot length is $L\sim150$ nm \cite{Deng:2018}, the magnetic field strengths over which a pinned ABS can be observed are roughly given by 0.5 T $\lesssim B\lesssim$ 3 T, consistent with experimental observations.

\section{Numerical Calculations}\label{numerical}
In this section, we perform numerical simulations of the hybrid system presented in Fig.~\ref{fig:Fig1}. We find that even when we incorporate the superconductor explicitly, all of the qualitative features of our analytical model from Sec.~\ref{analytical} are reproduced in the numerical results. Specifically, we show explicitly how the system can be fine tuned to a regime in which there is a low-energy ABS and demonstrate that this ABS is sensitive only to properties of the quantum dot. Finally, we show that if the quantum dot is removed from the system so no low-energy ABSs can form, no zero-energy states are present because the strong proximity coupling drives the topological phase transition to field strengths exceeding the critical field of the superconducting layer.

\begin{figure}[t!]
\includegraphics[width=\linewidth]{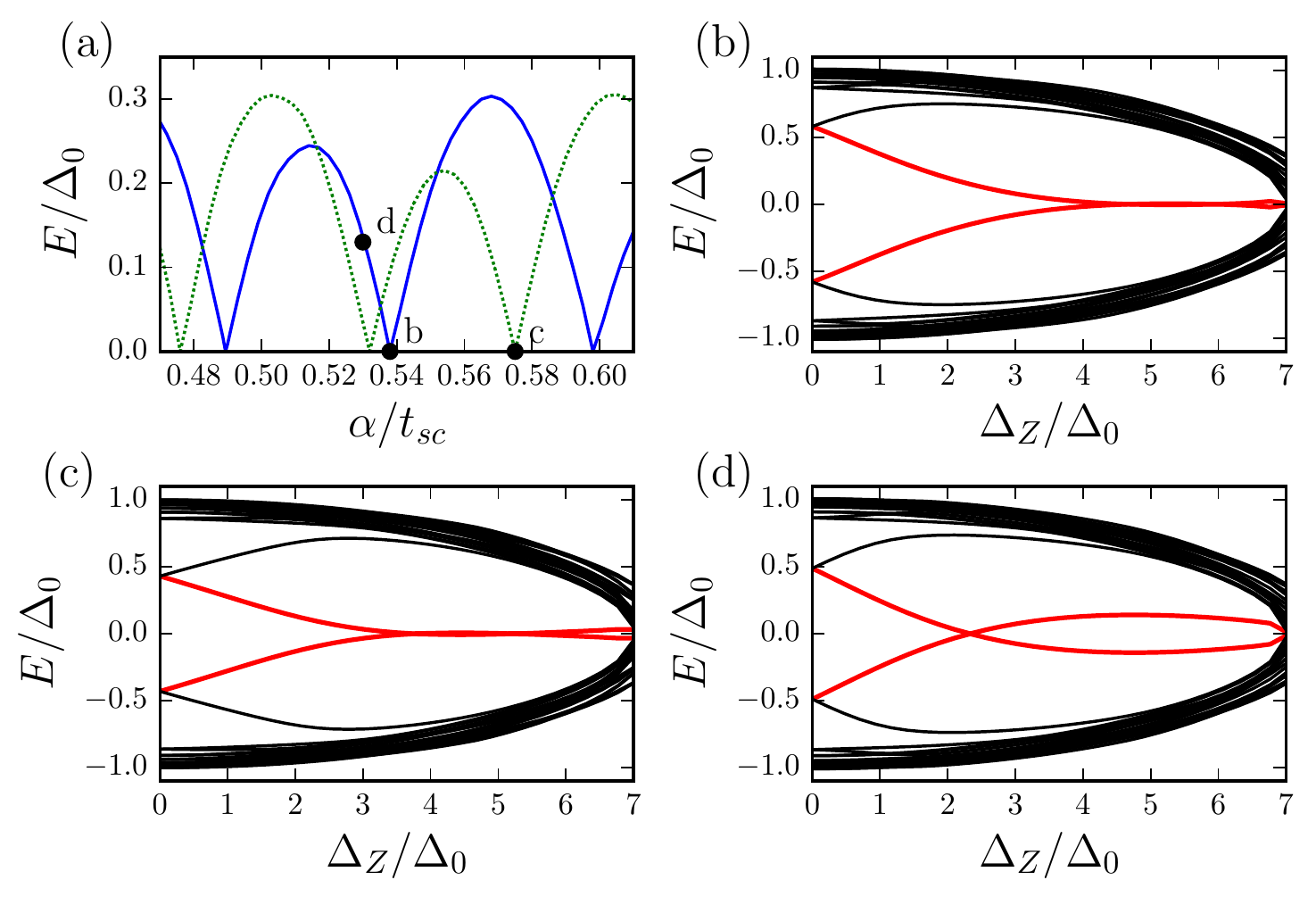}
\caption{(a) We first fix $\Delta_Z=0.75\Delta_Z^c$ and find that the energy $E$ of the lowest subgap ABS oscillates as a function of spin-orbit strength $\alpha$. The solid blue curve corresponds to $\mu=0$ while the dashed green curve corresponds to $\mu=-3\Delta_0$. The resonance condition, which is determined by points at which the energy of the ABS pass through zero, is altered by changes in $\mu$. (b) For $\mu=0$, we tune to the resonant value $\alpha/t_{sc}=0.538$ [as indicated in panel (a)] and find an ABS (red) that is pinned near zero energy over a large range of Zeeman energy $\Delta_Z$. Higher energy states are shown in black. (c) For $\mu=-3\Delta_0$, we tune to a different resonant value $\alpha/t_{sc}=0.575$ and again find a pinned ABS. (d) Tuning away from the resonant value of $\alpha$, we find that the ABS is no longer pinned to zero energy. All plots are obtained numerically  by diagonalizing Eq.~\eqref{Htb} for $L=300$, $L_s=2001$, $d=51$, $\mu_{sc}/t_{sc}=0.1$, $\Delta_0/t_{sc}=0.001$, $t_w/t_{sc}=5$, $t/t_{sc}=0.25$, and $\Delta_Z^c/\Delta_0=7$.
}
\label{fig2}
\end{figure}

We consider a discretized model described by
\begin{equation} \label{Htb}
H=H_w+H_s+H_t.
\end{equation}
The Hamiltonian of the nanowire, which we take to be a strictly 1D system with open boundaries, is given by
\begin{equation} \label{Hw}
\begin{aligned}
H_w&=\sum_{x=1}^{L+L_s-1}\biggl[b^\dagger_x(2t_w-\mu-\Delta_Z\sigma_1)b_x \\
	&\hspace{40pt} -\{b^\dagger_x(t_w-i\alpha\sigma_3/2)b_{x+1}+H.c.\}\biggr],
\end{aligned}
\end{equation}
where $b_x=(b_{x,\uparrow},b_{x,\downarrow})^T$ is a spinor of operators $b^\dagger_{x,\sigma}$ ($b_{x,\sigma}$) that create (destroy) an electron of spin $\sigma$ at site $x$ within the nanowire, which has length $L+L_s$, and $t_w=1/2m$ is the hopping amplitude (we set the lattice spacing $a=1$). [Note that $\alpha$ in Eq.~\eqref{Hw} is the same as that in Eq.~\eqref{model} due to the choice $a=1$.] The superconductor is described by a BCS Hamiltonian with constant $s$-wave pairing potential,
\begin{equation}
\begin{aligned}
H_{sc}&=\sum_{x=L+1}^{L+L_s-1}\sum_{y=1}^{d-1}\biggl[c_{x,y}^\dagger(4t_{sc}-\mu_{sc})c_{x,y}-\{t_{sc}c^\dagger_{x,y}c_{x+1,y} \\
	&+t_{sc}c^\dagger_{x,y}c_{x,y+1}+\Delta c^\dagger_{x,y,\downarrow}c^\dagger_{x,y,\uparrow}+H.c.\}\biggr],
\end{aligned}
\end{equation}
where $c_{x,y}=(c_{x,y,\uparrow},c_{x,y,\downarrow})^T$ is a spinor of operators $c^\dagger_{x,y,\sigma}$ ($c_{x,y,\sigma}$) that create (destroy) an electron of spin $\sigma$ at site $(x,y)$ within the superconductor. Again, the superconductor is taken to have open boundaries ($d$ is the thickness), and the magnetic field is incorporated only through the suppression of the pairing potential such that $\Delta=\Delta_0\sqrt{1-(\Delta_Z/\Delta_Z^c)^2}$. We note that $\Delta$ here is the pairing potential of the superconducting layer, whereas the $\Delta$ appearing in Sec.~\ref{analytical} represented the proximity-induced superconducting gap in the bulk of the nanowire. The nanowire and superconductor are coupled via local spin-conserving tunneling of the form
\begin{equation}
H_t=-t\sum_{x=L+1}^{L+L_s-1}\{c^\dagger_{x,1}b_x+H.c.\},
\end{equation}
where $t$ is the hopping amplitude between the nanowire and superconductor, which parametrizes the strength of the proximity coupling.

\begin{figure}[t!]
\includegraphics[width=\linewidth]{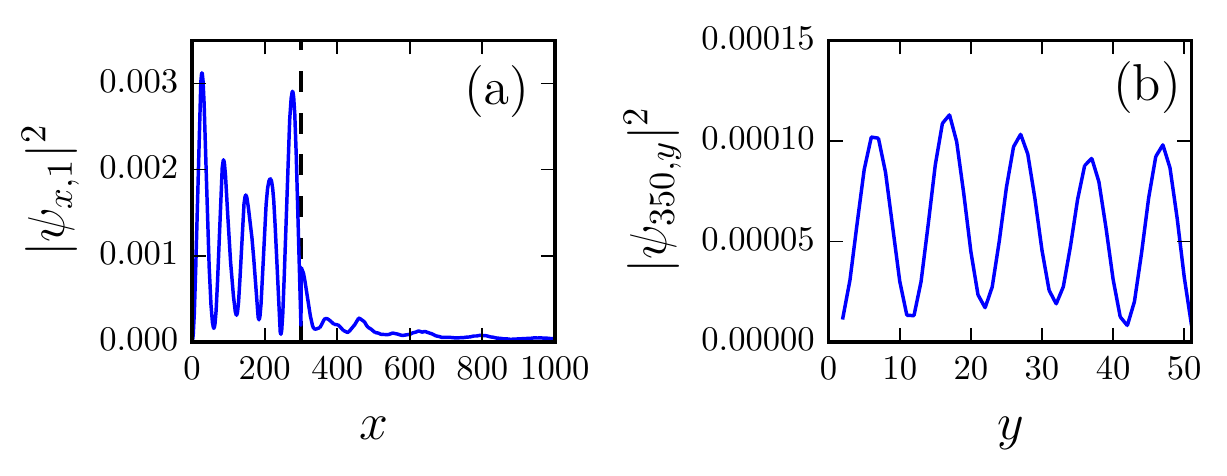}
\caption{\label{wf} Probability density $|\psi_{x,y}|^2$ of the low-energy ABS (a) within the nanowire ($y=1$) and (b) within the superconductor (with a line cut taken along $x=350$). Because $d$ is much smaller than the superconducting coherence length, the wave function does not decay in the $y$-direction. The total weight of the wave function  within the quantum dot is 0.32, the total weight in the proximitized nanowire region is 0.11, and the total weight in the superconductor is 0.57. Parameters are the same as in Fig.~\ref{fig2}(b), with $\Delta_Z/\Delta_0=5$.}
\end{figure}

To determine how to tune the system to observe an ABS pinned to zero energy, we first fix the magnetic field strength relatively close to the critical field of the superconductor and calculate the energy of the lowest subgap ABS as a function of spin-orbit strength $\alpha$. As the pinning behavior is optimized for $\mu=0$, we assume that the chemical potential is tuned in such a way. Consistent with the analytical model of Sec.~\ref{analytical} and as shown in Fig.~\ref{fig2}(a), the energy of the subgap ABS depends on $\alpha$ in an oscillatory way, with the energy passing through zero only for specific values of $\alpha$. After fixing $\alpha$ such that the ABS energy is small (i.e., after fine-tuning the system so there is a low-energy state), we then calculate the full ABS spectrum as a function of field strength. As shown in Fig.~\ref{fig2}(b), we find for this specific value of $\alpha$ that the ABS is pinned near zero energy for a large range of field strength \cite{spinx}. Note that no topological phase transition occurs in Fig.~\ref{fig2}(b), as this requires a field strength that exceeds the critical field of the superconductor, $B>B_c$ (we also show this explicitly in Appendix \ref{appB}). For the chosen parameters (which are given in the caption of Fig.~\ref{fig2}), we find that the ABS becomes pinned near zero energy at a Zeeman energy of about $\Delta_Z\approx4\Delta_0$ and remains pinned all the way up to $\Delta_Z^c$. This is also consistent with the analytical results of Sec.~\ref{analytical}, where it was shown that the ABS is pinned for Zeeman energies $\alpha/L\lesssim\Delta_Z\lesssim E_{so}$; for the parameters of Fig.~\ref{fig2}, these energy scales correspond to $\alpha/L\approx1.8\Delta_0$ and $E_{so}\approx2\Delta_Z^c$. Additionally, we show in Fig.~\ref{fig2}(c) that a pinned ABS can also be observed when the chemical potential is detuned away from $\mu=0$. We plot the probability density of the ABS within the nanowire in Fig.~\ref{wf}, showing that most of the weight within the nanowire is concentrated in the quantum dot; however, we also find that the wave function of the ABS is roughly spread equally between the nanowire and the superconducting layer.

\begin{figure}[t!]
\includegraphics[width=\linewidth]{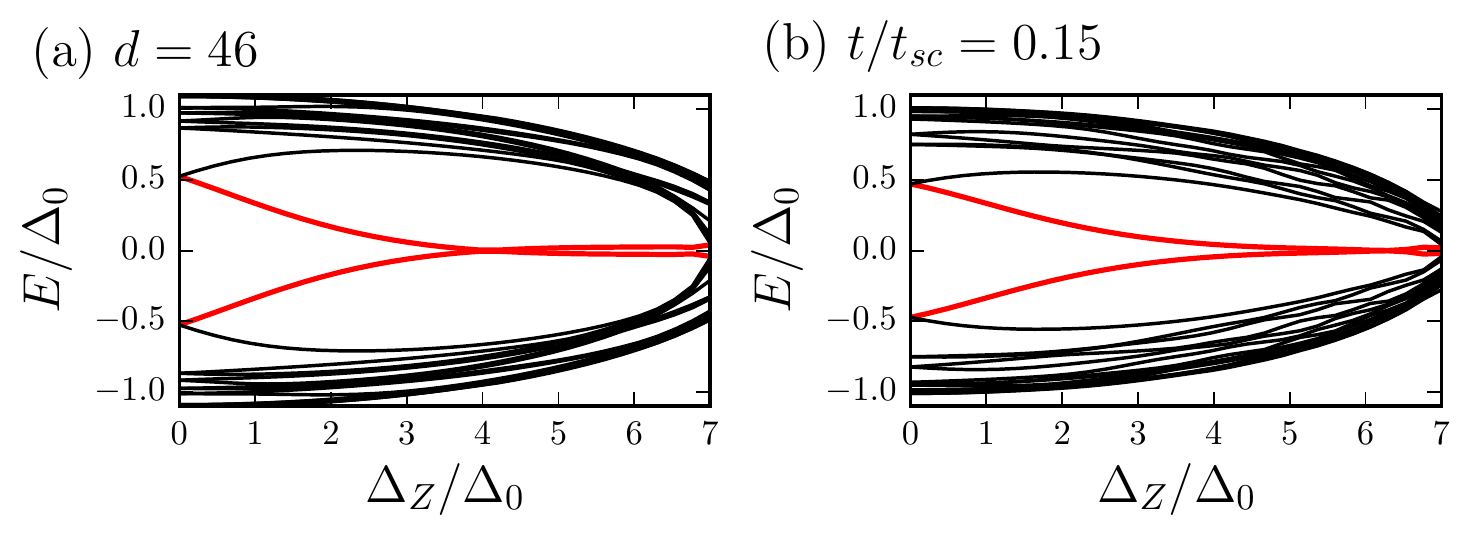}
\caption{Changing the thickness of the superconductor $d$ and the strength of the proximity coupling $t$ does not affect the pinning of the ABS despite significantly altering the properties of the nanowire region contacted by the superconductor, thus, indicating that the ABS is a property of the quantum dot. Parameters are the same as in Fig.~\ref{fig2}(b), with the exception $d=46$ in panel (a) and $t/t_{sc}=0.15$ in panel (b).
}
\label{fig3}
\end{figure}

To check the analytical result that the presence of a low-energy ABS is sensitive only to parameters of the normal dot region, we can test the robustness of the ABS to variations in both the thickness $d$ of the superconducting layer as well as the strength of the proximity coupling $t$ \cite{disorder}. It was shown in Refs.~\cite{Reeg:2017_3,Reeg:2018} that all properties of the region of nanowire coupled to the superconductor are highly dependent on both the thickness $d$ and the tunneling strength $t$. In particular, the effective chemical potential shift, and therefore the field strength at which a topological phase transition occurs, can vary greatly depending on the precise value of $d$, while the renormalization of material parameters is weakened as $t$ is made smaller. However, as shown in Fig.~\ref{fig3}, we find that varying the thickness $d$ [Fig.~\ref{fig3}(a)] and tunneling strength $t$ [Fig.~\ref{fig3}(b)] has no appreciable effect on the pinning of the ABS. As changing these parameters drastically alters the topological properties of the system (see Appendix~\ref{appB} for a more detailed discussion), we can conclude that the presence of a low-energy ABS is determined entirely by the properties of the dot region \cite{numerics}.

\begin{figure}[t!]
\centering
\includegraphics[width=.8\linewidth]{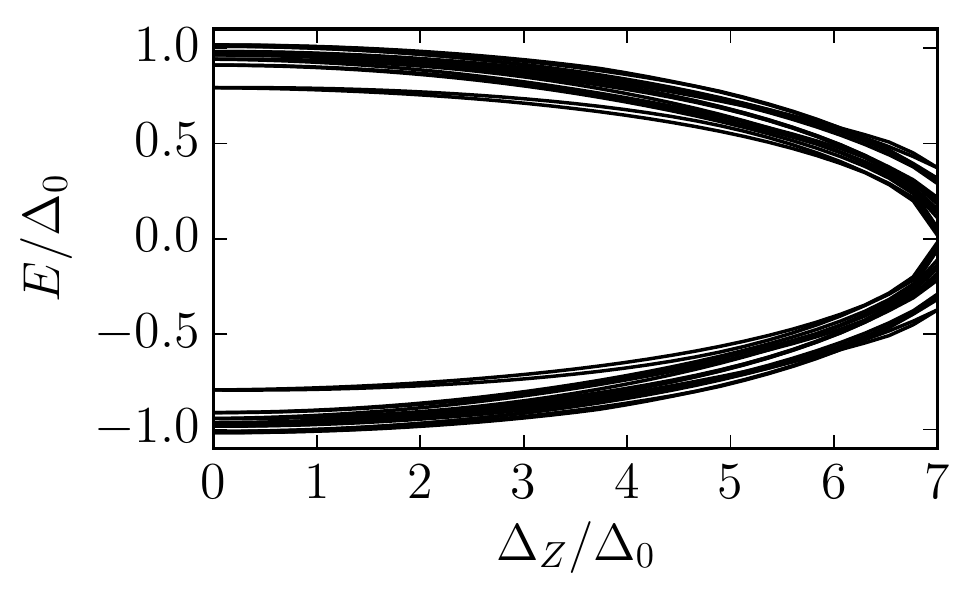}
\caption{Removing the quantum dot eliminates all low-energy ABSs from the system. No zero-energy states are observed because the topological phase transition corresponds to a magnetic field strength exceeding the critical field of the superconductor. Parameters are the same as in Fig.~\ref{fig2}(b), with the exception $L=0$. \label{nodot}
}
\label{fig4}
\end{figure}

In further support of this conclusion, we also calculate the spectrum of the system in the absence of a quantum dot ($L=0$). As shown in Fig.~\ref{nodot}, there are no low-energy states in this case because the topological phase transition occurs for magnetic field strength $B>B_c$. As a result, one essentially observes only the closing of the energy gap in the superconducting layer. Note that there is one subgap ABS, but this state remains very close to the gap edge, consistent with the findings of Refs.~\cite{Aseev:2018,Huang:2018}.

\section{Conclusions}\label{conclusion}
We studied a nanowire/superconductor hybrid system in which a region of the nanowire is left uncovered by the superconductor and forms a quantum dot, a geometry that has been studied extensively in many recent experiments \cite{Deng:2016,Zhang:2018,Vaitiekenas:2018,Deng:2018,deMoor:2018}. By considering a simple analytical model that incorporates the strong-coupling proximity effect \cite{Reeg:2017_3,Reeg:2018}, we showed that an ABS can be pinned exponentially close to zero energy, thus mimicking a topological zero-energy MBS, if the chemical potential is tuned within the Zeeman gap and if the spin-orbit length is tuned with respect to the dot length $L$ such that a resonance condition is satisfied. If this resonance condition is satisfied, the ABS will be pinned near zero energy for Zeeman energies $\alpha/L\lesssim\Delta_Z\lesssim E_{so}$, where $\alpha/L$ is the characteristic spacing between ABS levels and $E_{so}$ is the spin-orbit energy of the dot. Therefore, this ABS is purely a property of the quantum dot and unrelated to details of the proximity effect, and, for this reason, the physical mechanism giving rise to the ABS also applies to much more general cases than the experimentally relevant strong-coupling limit.

To supplement our analytical model, we also performed a numerical simulation of the hybrid system while explicitly incorporating the superconducting layer, thus going beyond the strict 1D models of Refs.~\cite{Liu:2017,Ptok:2017,Setiawan:2017,Moore:2018,Vuik:2018,Avila:2018}. We found that all of the qualitative features of our analytical model are also reproduced in the numerical results. Namely, we showed that if one first tunes the spin-orbit length such that a low-energy state is present at large magnetic field strength (corresponding to the resonance condition), one then generically finds that this ABS remains pinned close to zero energy over a large range of field strength. Additionally, we showed that the presence of this low-energy ABS is insensitive to both the thickness of the superconducting layer and the strength of the proximity effect, thus indicating that it is a property of the quantum dot rather than the proximitized region of the nanowire. Finally, we showed that if the quantum dot is removed, no low-energy ABSs are present in the system.

Our results demonstrate how easily one might be able to tune an experimental system (like that studied in Refs.~\cite{Deng:2016,Zhang:2018,Vaitiekenas:2018,Deng:2018,deMoor:2018}) to observe a trivial ABS pinned close to zero energy. As we have shown, if the voltages of the tunnel (contacting the quantum dot region) and back (contacting the nanowire) gates are tuned simultaneously (i.e., if the spin-orbit strength, quantum dot size, and chemical potential are tuned simultaneously) such that there is a low-energy state for magnetic field strengths near the critical field of the superconductor, one will generally then observe the pinning of a trivial ABS close to zero energy even in the absence of a topological phase transition, provided that the required separation of energy scales is satisfied. While we have obtained a sharp resonance condition by calculating the spectrum at zero temperature, it is worth noting that in a transport experiment this resonance can be broadened by finite temperature and the coupling to external leads \cite{Patrick,Diego,law,wimmer,denis2016,Nichele:2017}, thus increasing the likelihood of observing a pinned ABS. To avoid such low-energy trivial ABSs, it would be beneficial to remove the quantum dot from the system entirely \cite{Grivnin:2018}. The observation of a zero-energy state emerging as a function of magnetic field strength in this case would constitute a stronger local signature of a true topological MBS than a similar observation in the current experimental setup \cite{bulk}.

\emph{Acknowledgments.} This work was supported by the Swiss National Science Foundation and  NCCR QSIT. This project received funding from the European Union's Horizon 2020 research and innovation program (ERC Starting Grant, grant agreement No 757725).

\appendix
\section{Numerical solution of effective 1D model} \label{appA}

\begin{figure}[t!]
\includegraphics[width=\linewidth]{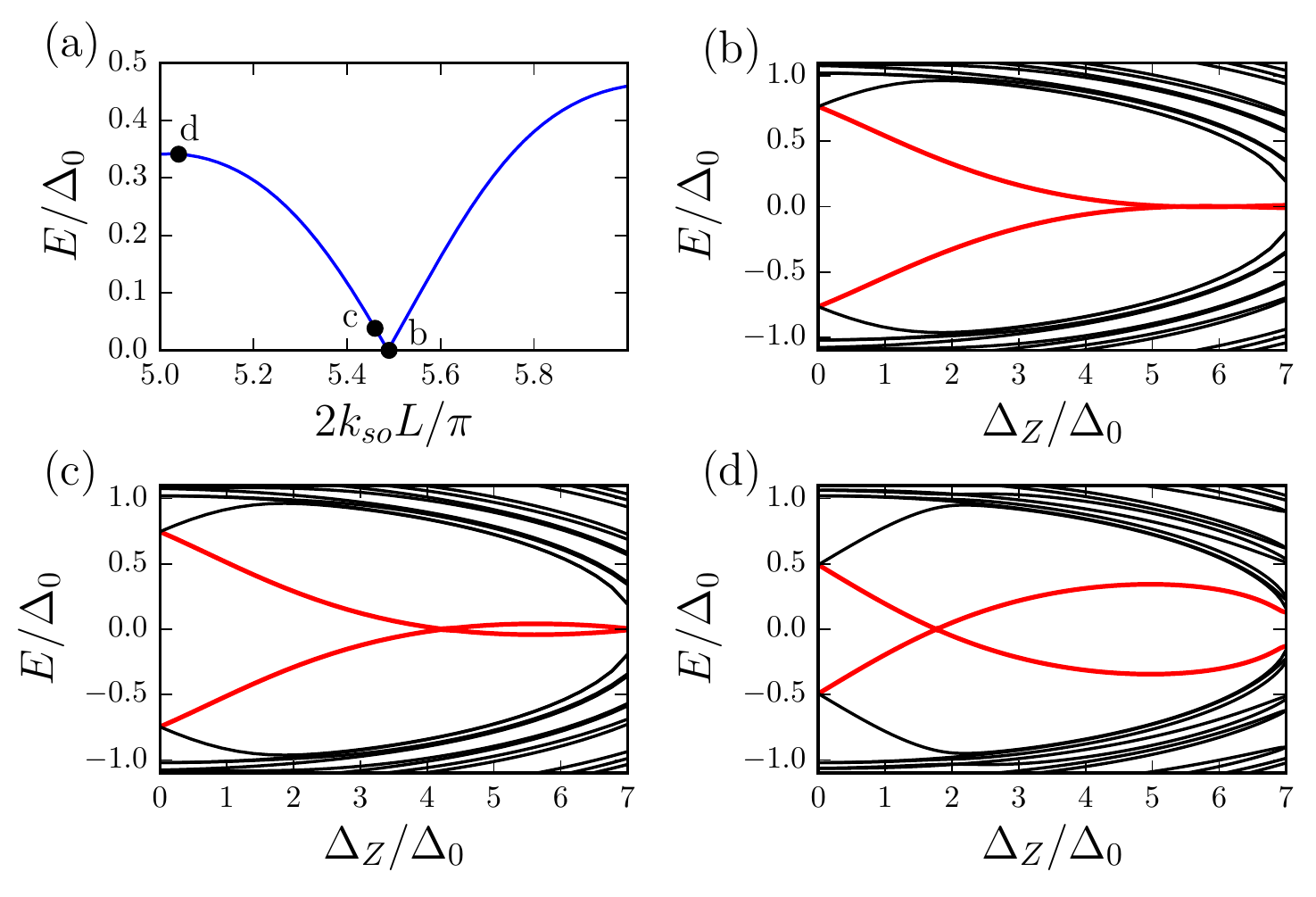}
\caption{(a) At $\Delta_Z=0.75\Delta_Z^c$, the energy $E$ of the lowest subgap ABS varies as a function of $k_{so}L$. Consistent with the resonance condition found analytically in Sec.~\ref{analytical}, the ABS energy becomes small near $\cos(2k_{so}L)=0$. (b) Tuning to the resonant value $2k_{so}L/\pi=5.49$, we find an ABS (red) that is pinned near zero energy over a large range of Zeeman energy $\Delta_Z$. (c-d) Away from the resonant value [we take $2k_{so}L/\pi=5.46$ in panel (c) and $2k_{so}L/\pi=5.04$ in panel (d)], the ABS is no longer pinned near zero energy. All plots are obtained by numerically diagonalizing the Hamiltonian of Eq.~\eqref{tight-binding} for $L=300$ nm, $L_s=2$ $\mu$m, $\mu_s=2$ meV, $\mu=0$, $\Delta_0=0.25$ meV, $t_s=20$ meV, $t_w = 100$ meV, and $\Delta_Z^c=1.75$ meV. Note that $2k_{so}L/\pi=5.49$ corresponds to a spin-orbit energy $E_{so}=2.06$ meV, and that the lattice constant is taken to be $a=5$ nm so that $L=60a$ and $L_s=400a$.}
\label{1Dmodel}
\end{figure}

In this Appendix, we numerically solve the minimal analytical model defined by $H_{\rm eff}$ in Eq.~\eqref{model}. Crucially, we demonstrate that the strong inequalities determining the ABS pinning obtained analytically, $\alpha/L\ll\Delta_Z\ll E_{so}$ can be relaxed to $\alpha/L\lesssim\Delta_Z\lesssim E_{so}$ while retaining the same qualitative behavior. We again stress that the model considered here is distinct from that considered in Sec.~\ref{numerical} and is not restricted to describe a hybrid device in the strong-coupling regime.

The discretized version of the Hamiltonian $H_{\rm eff}$ in Eq.~\eqref{model} reads
\begin{equation}  \label{tight-binding}
\begin{aligned}
&\bar H_{\rm eff}= \sum_{x=1}^{L+L_s-1}\biggl[c^\dagger_x (t_{w,x+1/2}+t_{w,x-1/2}-\mu_x \\
	&-\Delta_{Z,x}\sigma_1)c_x  -\{c^\dagger_x (t_{w,x+1/2} \\
	&-i \alpha_{x+1/2} \sigma_3/2 )c_{x+1}+\Delta_x c^\dagger_{x,\downarrow}c^\dagger_{x,\uparrow}+H.c.\}\biggr], 
\end{aligned}
\end{equation}
where $c_x=(c_{x,\uparrow},c_{x,\downarrow})^T$ is a spinor of operators $c^\dagger_{x,\sigma}$ ($c_{x,\sigma}$) that create (destroy) an electron of spin $\sigma$ at site $x$, and we again set the lattice constant $a=1$. As in Eq.~\eqref{model}, we assume that all parameters take different values in the normal and superconducting regions of the nanowire, where the boundary is taken at $L_b=L+1/2$: $\alpha_x=\alpha\theta(L_b-x)$, $\Delta_{Z,x}=\Delta_Z\theta(L_b-x)$, $t_{w,x}=t_w\theta(L_b-x)+t_s\theta(x-L_b)$, $\Delta_x=\Delta_0\sqrt{1-(\Delta_Z/\Delta_Z^c)^2}\theta(x-L_b)$, and $\mu_x=\mu\theta(L_b-x)+\mu_s\theta(x-L_b)$. Note that the Heaviside step function is defined such that $\theta(0)=1/2$.

\begin{figure}[t!]
\includegraphics[width=\linewidth]{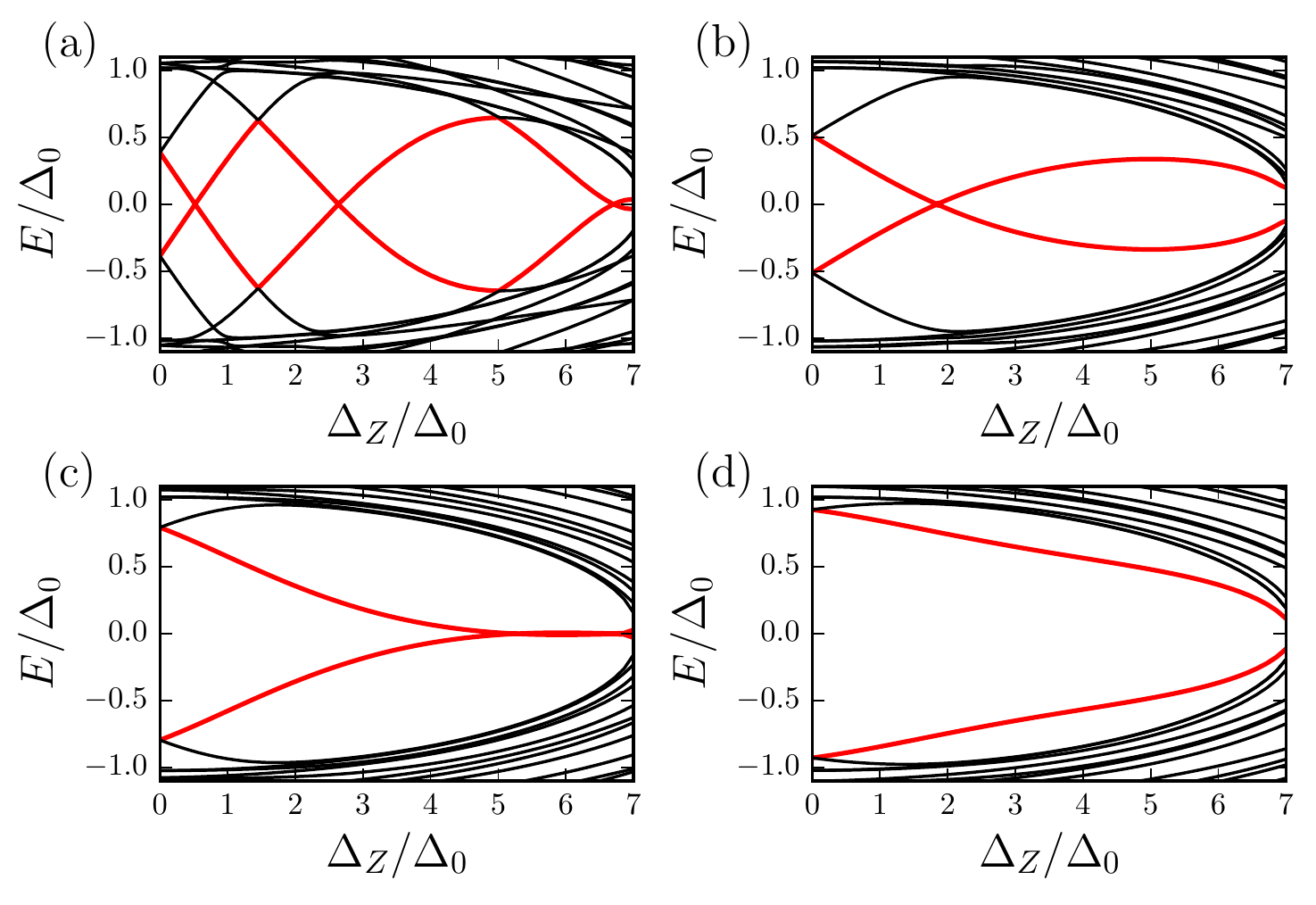}
\caption{(a) Energy spectrum found by diagonalizing the Hamiltonian in Eq.~\eqref{tight-binding} in the absence of the SOI ($\alpha =0$), with the lowest energy ABS shown in red. (b) For $\alpha/t_w=0.26$, an ABS crosses zero energy for a single value of $\Delta_Z$. (c) For $\alpha/t_w=0.2827$, an ABS is pinned near zero energy over a large range of $\Delta_Z$. (d) For $\alpha/t_w=0.31$, an ABS stays away from zero energy for all values of $\Delta_Z$. All remaining parameters are the same as in Fig.~\ref{1Dmodel}, with the exception $\mu=0.1$ meV.
}
\label{1Dmodel2}
\end{figure}

As shown in Fig.~\ref{1Dmodel}(a), we are able to reproduce numerically the resonance condition $\cos(2k_{so}L)=0$ (for the case $\mu=0$) that was found analytically in Sec.~\ref{analytical}. When the SOI strength is tuned to its resonant value, we obtain an ABS that is pinned near zero energy over a large range of Zeeman energy $\Delta_Z$ [Fig.~\ref{1Dmodel}(b)].  We note that when the resonance condition is satisfied for the parameters of Fig.~\ref{1Dmodel}(b), the ABS is pinned between $\Delta_Z\approx4\Delta_0$ and $\Delta_Z\approx7\Delta_0$. Comparing with $\alpha/L=1.9\Delta_0$ and $E_{so}=8.3\Delta_0$, we see that the strong inequalities obtained analytically to describe the range of Zeeman energy over which the ABS is pinned can be relaxed to weaker inequalities $\alpha/L\lesssim\Delta_Z\lesssim E_{so}$. This is also consistent with the numerical results of Sec.~\ref{numerical}. However, note that because the energy range $\alpha/L$ to $E_{so}$ is rather small, a pinned ABS is present only for the single resonant value shown in Fig.~\ref{1Dmodel}(a). Away from resonance, the ABS is no longer pinned near zero energy, as shown in Figs.~\ref{1Dmodel}(c-d). Next, we show in Fig.~\ref{1Dmodel2} that if the chemical potential is detuned away from $\mu=0$, a new resonance condition must be satisfied to see a pinned ABS, consistent with the numerical results presented in Fig.~\ref{fig2}(c).

\begin{figure}[b!]
\includegraphics[width=0.8\linewidth]{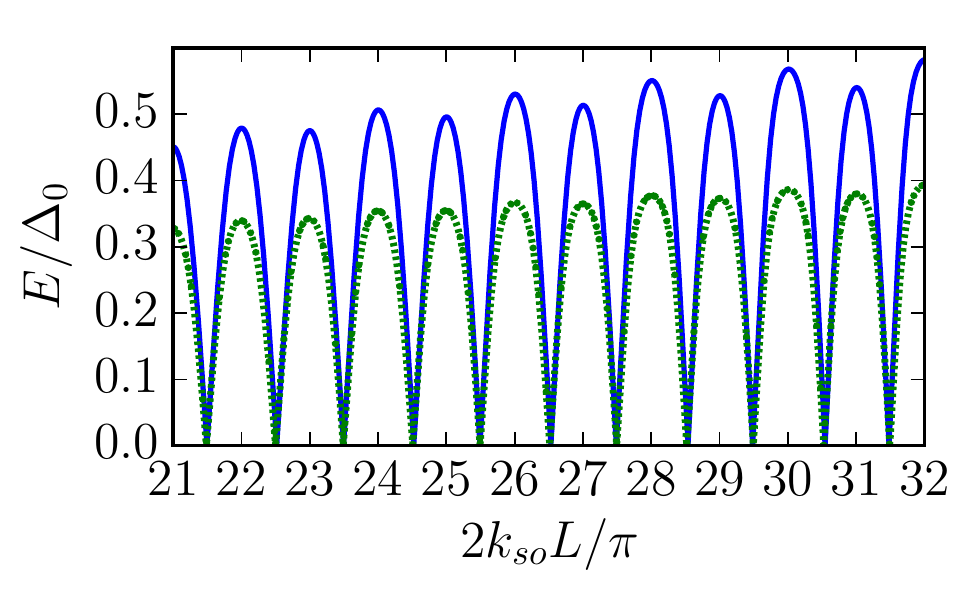}
\caption{\label{evsalarge} The energy $E$ of the ABS goes to zero for both $\Delta_Z=0.75\Delta_Z^c$ (solid blue curve) and $\Delta_Z=0.9\Delta_Z^c$ (dashed green curve) at the same values of $k_{so}L$, indicating a pinned ABS at all resonant values over the range of the plot. Furthermore, the resonant values occur when $\cos(2k_{so}L)=0$, consistent with the analytical results of Sec.~\ref{analytical}. This plot was obtained by numerically diagonalizing Eq.~\eqref{tight-binding} with $L=750$ nm, $L_s=2$ $\mu$m, $\mu_s=2$ meV, $\mu=0$, $\Delta_0=0.25$ meV, $t_s=2$ eV, $ t_w=10$ eV, and $\Delta_Z^c=1.75$ meV. Note that the lattice constant is taken to be $a=0.5$ nm so that $L=1500a$ and $L_s=4000a$.}
\end{figure}

While we showed in Fig.~\ref{1Dmodel} that a pinned ABS can be obtained for $\alpha/L\lesssim\Delta_Z\lesssim E_{so}$, this was true only for a single value of $2k_{so}L/\pi$. Based on the analytical calculation presented in Sec.~\ref{analytical}, it should be possible to have a pinned ABS for additional values of $2k_{so}L/\pi$ if the energy range $\alpha/L$ to $E_{so}$ is expanded. We show that this statement holds numerically in Fig.~\ref{evsalarge}, where we plot the energy of the ABS as a function of $k_{so}L$ for the Zeeman energies $\Delta_Z=0.75\Delta_Z^c$ and $\Delta_Z=0.9\Delta_Z^c$. Because the ABS energy goes to zero for $2k_{so}L/\pi\in(\mathbb{Z}+1/2)$ for both values of $\Delta_Z$ over the whole range of the plot, a pinned ABS exists at each resonant value. This confirms the analytical statement that a pinned ABS occurs when $\cos(2k_{so}L)=0$ provided that $\alpha/L\ll\Delta_Z\ll E_{so}$.

\begin{figure}[t!]
\includegraphics[width=0.8\linewidth]{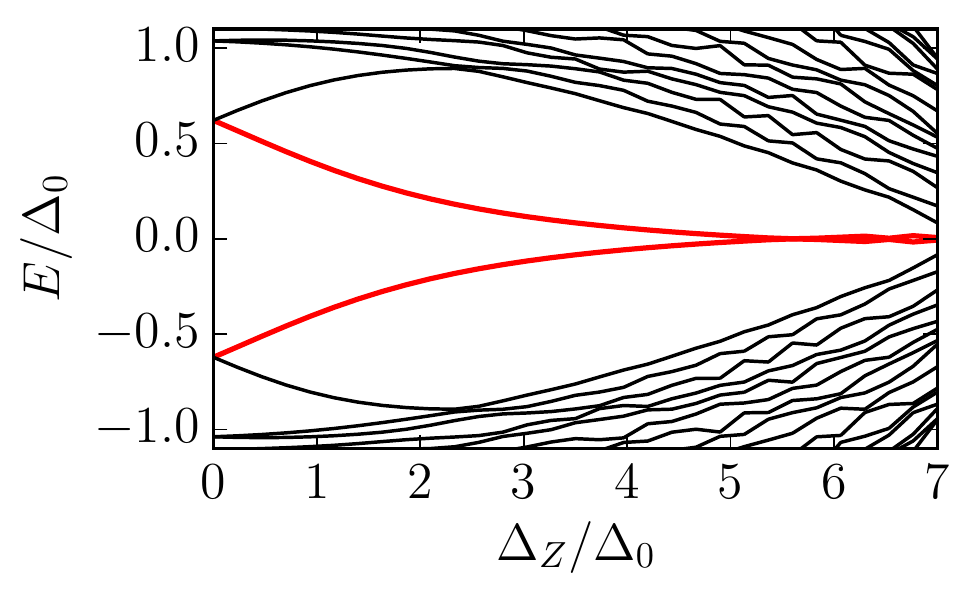}
\caption{\label{uniform} When material parameters are spatially uniform, it is also possible to obtain a pinned ABS. This plot is obtained by diagonalizing Eq.~\eqref{tight-binding} with $\alpha_x=\alpha$, $\Delta_{Z,x}=\Delta_Z$, and $t_{w,x}=t_w$. We additionally choose $L=500$ nm, $L_s=5$ $\mu$m, $\mu_s=2$ meV, $\mu=0$, $\Delta_0=0.25$ meV, $t_w=100$ meV, $\Delta_Z^c=1.75$ meV, and $E_{so}=2.13$ meV. The lattice constant is taken as $a=5$ nm so that $L=100a$ and $L_s=1000a$.}
\end{figure}

While we have to this point considered the case where the superconducting region does not have SOI or Zeeman energy (corresponding strictly to the analytical model considered in Sec.~\ref{analytical}), we also noted previously that the analytical calculation is valid for any general case for which $\mu_s\gg E_{so,s},\Delta_{Z,s}$. To illustrate this numerically, let us consider the Hamiltonian of Eq.~\eqref{tight-binding} assuming that the SOI, Zeeman energy, and hopping amplitude are uniform throughout the nanowire; i.e., $\alpha_x=\alpha$, $\Delta_{Z,x}=\Delta_Z$, and $t_{w,x}=t_w$, while we keep $\Delta_x$ and $\mu_x$ as defined below Eq.~\eqref{tight-binding}. As shown in Fig.~\ref{uniform}, we also find a pinned ABS in this case; note that no topological transition occurs in Fig.~\ref{uniform} because $\mu_s>\Delta_Z^c$.

\begin{figure*}[t!]
\includegraphics[width=\linewidth]{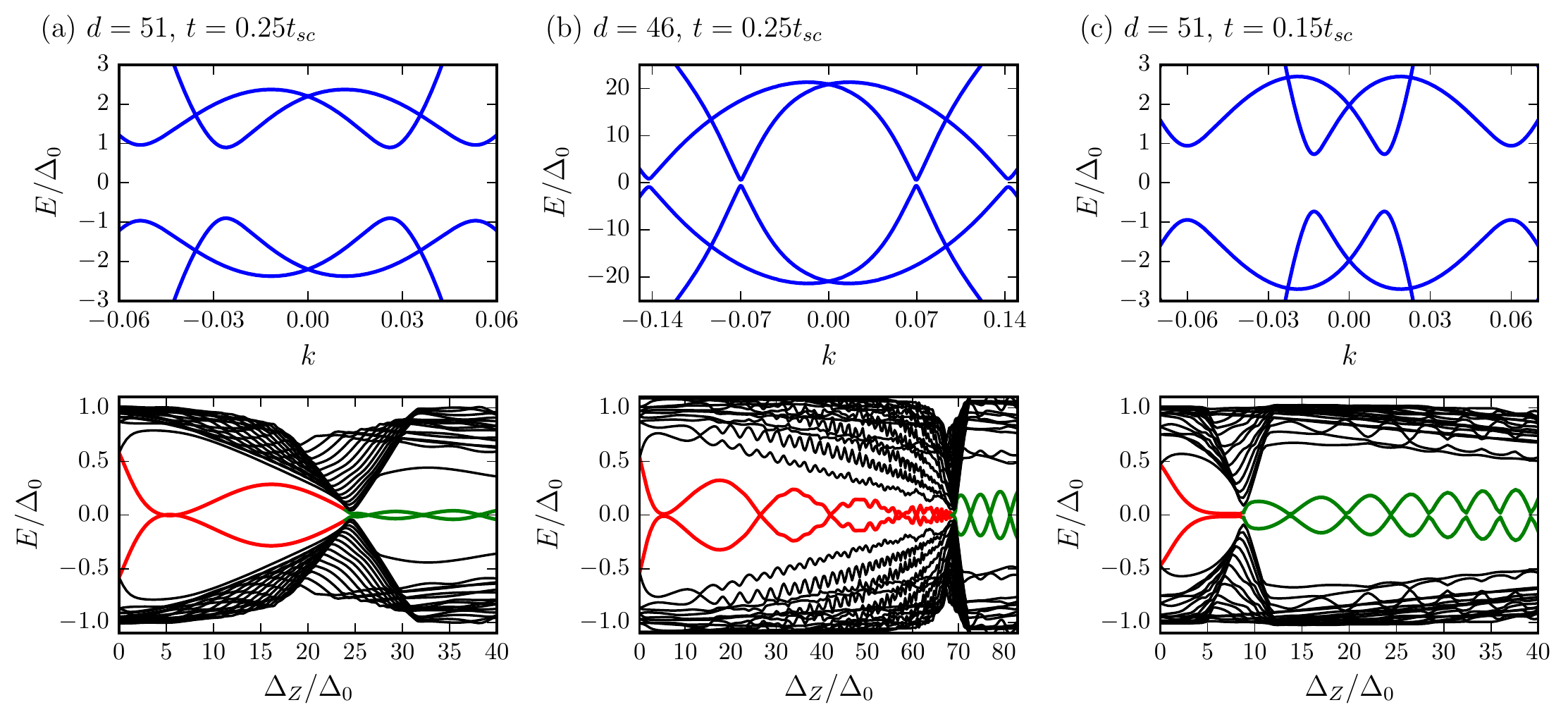}
\caption{Top panels: Bulk spectra obtained by numerically diagonalizing Eq.~\eqref{Htbk} for an infinite system with no quantum dot and for $\Delta_Z=0$. Note that we plot only the lowest subband of the spectrum in each case. Bottom panels: Spectra obtained by numerically diagonalizing Eq.~\eqref{Htb} in a finite system (with a quantum dot). To access large magnetic fields, we take $B_c\to\infty$. (a) Top: In the absence of a magnetic field, we find that the proximity effect significantly reduces the spin-orbit energy of the nanowire and induces an effective chemical potential shift. Bottom: The evolution of the spectrum with Zeeman energy reveals a topological phase transition at $\Delta_Z\approx25\Delta_0$, beyond which there is a topological MBS (green). The pinning of the ABS (red) near zero energy observed in Fig.~\ref{fig2}(b) is also seen here between $\Delta_Z\approx4\Delta_0$ and $\Delta_Z\approx 7\Delta_0$. (b) Top: Changing the superconductor thickness $d$ leads to significant change in the effective chemical potential of the proximitized nanowire. Bottom: Due to a larger effective chemical potential, the topological phase transition is shifted to much higher Zeeman energy $\Delta_Z\approx70\Delta_0$. However, the behavior of the trivial ABS remains unchanged. (c) Top: Reducing the tunneling strength $t$ leads to approximately the same chemical potential shift, but the renormalization of the spin-orbit energy is also reduced. Bottom: For weaker coupling $t$, the $g$-factor of the proximitized nanowire is larger (\emph{i.e.}, it is less renormalized) and a topological phase transition occurs at smaller Zeeman energy $\Delta_Z\approx8\Delta_0$. Again, the behavior of the ABS before the topological transition remains unchanged. All plots are obtained for $\mu_{sc}/t_{sc}=0.1$, $\mu=0$, $\alpha/t_{sc}=0.538$, $\Delta_0/t_{sc}=0.001$, $t_w/t_{sc}=5$, and $\Delta_Z^c\to\infty$. Plots in the bottom panels additionally have $L=300$ and $L_s=2000$. The superconductor thickness $d$ and tunneling strength $t$ of each panel are specified in the figure [(a) $d=51$, $t/t_{sc}=0.25$, (b) $d=46$, $t/t_{sc}=0.25$, and (c) $d=51$, $t/t_{sc}=0.15$].
 \label{bulk}}
\end{figure*}

\section{Bulk spectrum and topological \\ phase transition}\label{appB}
In Sec.~\ref{numerical} of the main text, we argued that the insensitivity of the pinning of the subgap ABS to variations in the superconductor thickness $d$ and tunneling strength $t$ indicate that the ABS is a property of the quantum dot rather than the proximitized region of the nanowire. In this appendix, we support this claim by showing explicitly how the topological properties of the system depend on both the thickness $d$ and tunneling strength $t$. This was studied extensively in Refs.~\cite{Reeg:2017_3,Reeg:2018} but we include a brief overview here for completeness.

To study how the proximitized region of the nanowire is affected by the strong proximity coupling, we can examine the bulk spectrum of this region. To access the bulk spectrum numerically, we assume that both the nanowire and superconductor are infinitely long (and therefore that there is no quantum dot region). We can then define a conserved momentum $k$ along the nanowire axis and describe the system by the Hamiltonian
\begin{equation} \label{Htbk}
\begin{aligned}
H&=\sum_{k}\biggl\{b_k^\dagger[2t_w(1-\cos k)-\mu-\Delta_Z\sigma_1-\alpha\sin k\sigma_3]b_k \\
	&+\sum_{y=1}^{d-1}\bigg[c^\dagger_{k,y}[2t_{sc}(2-\cos k)-\mu_{sc}]c_{k,y}-[t_{sc}c^\dagger_{k,y}c_{k,y+1} \\
	&+\Delta c^\dagger_{-k,y,\downarrow}c^\dagger_{k,y,\uparrow}+H.c.\biggr]-t[c^\dagger_{k,1}b_k+H.c.]\biggr\},
\end{aligned}
\end{equation}
which is simply the Fourier transformed version of Eq.~\eqref{Htb}.

First, we consider the parameters used to generate Fig.~\ref{fig2}(b), with a superconductor thickness $d=51$ sites and a tunneling strength $t=0.25t_{sc}$. The bulk spectrum for this case in the absence of a magnetic field ($\Delta_Z=0$) is plotted in the top panel of Fig.~\ref{bulk}(a). Due to the strong proximity coupling, the spin-orbit energy is significantly renormalized; for $t=0$ the spin-orbit energy is $E_{so}=14.5\Delta_0$, while for $t=0.25t_{sc}$ the spin-orbit energy is only $E_{so}=0.2\Delta_0$. Additionally, the proximity coupling induces an effective chemical potential shift that is comparable to the induced superconducting gap. The combination of a chemical potential shift and strong renormalization of the nanowire $g$-factor pushes the topological phase transition to very high magnetic field strength. This can be seen in the bottom panel of Fig.~\ref{bulk}(a), where we calculate the spectrum of a finite nanowire with a quantum dot and set $B_c\to\infty$ to allow for arbitrarily large fields without destroying superconductivity (while this is certainly unrealistic physically, we do so only to determine the field corresponding to the topological transition). We see that the topological transition in this case occurs near $\Delta_Z\approx25\Delta_0$, and we also see the pinned ABS between $\Delta_Z\approx4\Delta_0$ and $\Delta_Z\approx7\Delta_0$ that appeared in Fig.~\ref{fig2}(b).

Changing the thickness of the superconducting layer to $d=46$ sites induces a substantially larger effective chemical potential shift, as can be seen in the top panel of Fig.~\ref{bulk}(b). Due to the much larger effective chemical potential, the topological transition is then pushed to significantly higher magnetic field strength ($\Delta_Z\approx70\Delta_0$), as can be seen in the bottom panel of Fig.~\ref{bulk}(b). However, despite significant changes to the proximitized region of the nanowire, the behavior of the trivial ABS that originates in the quantum dot region remains essentially unaffected by the change in thickness $d$.

Reducing the tunneling strength between the nanowire and superconductor to $t=0.15t_{sc}$ leads to roughly the same chemical potential shift, but the parameters of the nanowire are significantly less renormalized by the proximity effect, as can be seen by examining the spin-orbit splitting of the spectrum in the top panel of Fig.~\ref{bulk}(c). Due to a larger $g$-factor, the topological phase transition is then pushed to smaller Zeeman energy ($\Delta_Z\approx8\Delta_0$) in the weaker proximity case, as shown in the bottom panel of Fig.~\ref{bulk}(c). Again, though, the behavior of the trivial ABS (before the topological transition) is unaffected by the change in tunneling strength.

Beyond the topological phase transitions in the bottom panels of Fig.~\ref{bulk}, a topological MBS emerges at each end of the system (shown in green). Due to the finite length of the nanowire and the large magnetic field strength required to reach the topological phase, the two MBSs have significant overlap and, as a result, the MBS energy is finite and oscillates as a function of $\Delta_Z$ \cite{Diego,madrid}. Comparing, for example, the bottom panels of Fig.~\ref{bulk}(b) and Fig.~\ref{bulk}(c), we see that it could be difficult to distinguish whether a bound state whose energy oscillates as a function of $\Delta_Z$ is an ABS or a MBS. From Fig.~\ref{bulk}, however, we find two distinguishing features. First, the period of the MBS oscillations is significantly shorter than the period of the ABS oscillations, as can be seen in all three panels of Fig.~\ref{bulk}. Second, the amplitude of the ABS oscillations decreases as a function of $\Delta_Z$ [seen clearly in Fig.~\ref{bulk}(b)] while the amplitude of the MBS oscillations increases as a function of $\Delta_Z$ [seen clearly in Fig.~\ref{bulk}(c)].

In this Appendix, we have demonstrated that while the topological properties of the nanowire/superconductor hybrid system (and by extension the properties of the proximitized region of the nanowire) are highly dependent on both the superconductor thickness $d$ and the tunneling strength $t$, the pinning of the low-energy ABS (between $\Delta_Z\approx4\Delta_0$ and $\Delta_Z\approx7\Delta_0$) is unaffected by these two parameters. As discussed in the main text, this is very strong evidence that the low-energy pinning of the ABS is related only to the quantum dot region and is unrelated to any topological properties of the system.

\bibliography{draft_self}

\begin{thebibliography}{83}%
\makeatletter
\providecommand \@ifxundefined [1]{%
 \@ifx{#1\undefined}
}%
\providecommand \@ifnum [1]{%
 \ifnum #1\expandafter \@firstoftwo
 \else \expandafter \@secondoftwo
 \fi
}%
\providecommand \@ifx [1]{%
 \ifx #1\expandafter \@firstoftwo
 \else \expandafter \@secondoftwo
 \fi
}%
\providecommand \natexlab [1]{#1}%
\providecommand \enquote  [1]{``#1''}%
\providecommand \bibnamefont  [1]{#1}%
\providecommand \bibfnamefont [1]{#1}%
\providecommand \citenamefont [1]{#1}%
\providecommand \href@noop [0]{\@secondoftwo}%
\providecommand \href [0]{\begingroup \@sanitize@url \@href}%
\providecommand \@href[1]{\@@startlink{#1}\@@href}%
\providecommand \@@href[1]{\endgroup#1\@@endlink}%
\providecommand \@sanitize@url [0]{\catcode `\\12\catcode `\$12\catcode
  `\&12\catcode `\#12\catcode `\^12\catcode `\_12\catcode `\%12\relax}%
\providecommand \@@startlink[1]{}%
\providecommand \@@endlink[0]{}%
\providecommand \url  [0]{\begingroup\@sanitize@url \@url }%
\providecommand \@url [1]{\endgroup\@href {#1}{\urlprefix }}%
\providecommand \urlprefix  [0]{URL }%
\providecommand \Eprint [0]{\href }%
\providecommand \doibase [0]{http://dx.doi.org/}%
\providecommand \selectlanguage [0]{\@gobble}%
\providecommand \bibinfo  [0]{\@secondoftwo}%
\providecommand \bibfield  [0]{\@secondoftwo}%
\providecommand \translation [1]{[#1]}%
\providecommand \BibitemOpen [0]{}%
\providecommand \bibitemStop [0]{}%
\providecommand \bibitemNoStop [0]{.\EOS\space}%
\providecommand \EOS [0]{\spacefactor3000\relax}%
\providecommand \BibitemShut  [1]{\csname bibitem#1\endcsname}%
\let\auto@bib@innerbib\@empty
\bibitem [{\citenamefont {Kitaev}(2001)}]{Kitaev:2001}%
  \BibitemOpen
  \bibfield  {author} {\bibinfo {author} {\bibfnamefont {A.~Y.}\ \bibnamefont
  {Kitaev}},\ }\href {http://stacks.iop.org/1063-7869/44/i=10S/a=S29}
  {\bibfield  {journal} {\bibinfo  {journal} {Physics-Uspekhi}\ }\textbf
  {\bibinfo {volume} {{\bf 44}}},\ \bibinfo {pages} {131} (\bibinfo {year}
  {2001})}\BibitemShut {NoStop}%
\bibitem [{\citenamefont {Alicea}(2012)}]{Alicea:2012}%
  \BibitemOpen
  \bibfield  {author} {\bibinfo {author} {\bibfnamefont {J.}~\bibnamefont
  {Alicea}},\ }\href {http://stacks.iop.org/0034-4885/75/i=7/a=076501}
  {\bibfield  {journal} {\bibinfo  {journal} {Rep. Prog. Phys.}\ }\textbf
  {\bibinfo {volume} {{\bf 75}}},\ \bibinfo {pages} {076501} (\bibinfo {year}
  {2012})}\BibitemShut {NoStop}%
\bibitem [{\citenamefont {Beenakker}(2013)}]{Beenakker:2013}%
  \BibitemOpen
  \bibfield  {author} {\bibinfo {author} {\bibfnamefont {C.~W.~J.}\
  \bibnamefont {Beenakker}},\ }\href
  {http://www.annualreviews.org/doi/abs/10.1146/annurev-conmatphys-030212-184337}
  {\bibfield  {journal} {\bibinfo  {journal} {Annu. Rev. Condens. Matter
  Phys.}\ }\textbf {\bibinfo {volume} {{\bf 4}}},\ \bibinfo {pages} {113}
  (\bibinfo {year} {2013})}\BibitemShut {NoStop}%
\bibitem [{\citenamefont {Oreg}\ \emph {et~al.}(2010)\citenamefont {Oreg},
  \citenamefont {Refael},\ and\ \citenamefont {von Oppen}}]{Oreg:2010}%
  \BibitemOpen
  \bibfield  {author} {\bibinfo {author} {\bibfnamefont {Y.}~\bibnamefont
  {Oreg}}, \bibinfo {author} {\bibfnamefont {G.}~\bibnamefont {Refael}}, \ and\
  \bibinfo {author} {\bibfnamefont {F.}~\bibnamefont {von Oppen}},\ }\href
  {\doibase 10.1103/PhysRevLett.105.177002} {\bibfield  {journal} {\bibinfo
  {journal} {Phys. Rev. Lett.}\ }\textbf {\bibinfo {volume} {{\bf 105}}},\
  \bibinfo {pages} {177002} (\bibinfo {year} {2010})}\BibitemShut {NoStop}%
\bibitem [{\citenamefont {Lutchyn}\ \emph {et~al.}(2010)\citenamefont
  {Lutchyn}, \citenamefont {Sau},\ and\ \citenamefont
  {Das~Sarma}}]{Lutchyn:2010}%
  \BibitemOpen
  \bibfield  {author} {\bibinfo {author} {\bibfnamefont {R.~M.}\ \bibnamefont
  {Lutchyn}}, \bibinfo {author} {\bibfnamefont {J.~D.}\ \bibnamefont {Sau}}, \
  and\ \bibinfo {author} {\bibfnamefont {S.}~\bibnamefont {Das~Sarma}},\ }\href
  {\doibase 10.1103/PhysRevLett.105.077001} {\bibfield  {journal} {\bibinfo
  {journal} {Phys. Rev. Lett.}\ }\textbf {\bibinfo {volume} {{\bf 105}}},\
  \bibinfo {pages} {077001} (\bibinfo {year} {2010})}\BibitemShut {NoStop}%
\bibitem [{\citenamefont {Mourik}\ \emph {et~al.}(2012)\citenamefont {Mourik},
  \citenamefont {Zuo}, \citenamefont {Frolov}, \citenamefont {Plissard},
  \citenamefont {Bakkers},\ and\ \citenamefont {Kouwenhoven}}]{Mourik:2012}%
  \BibitemOpen
  \bibfield  {author} {\bibinfo {author} {\bibfnamefont {V.}~\bibnamefont
  {Mourik}}, \bibinfo {author} {\bibfnamefont {K.}~\bibnamefont {Zuo}},
  \bibinfo {author} {\bibfnamefont {S.~M.}\ \bibnamefont {Frolov}}, \bibinfo
  {author} {\bibfnamefont {S.~R.}\ \bibnamefont {Plissard}}, \bibinfo {author}
  {\bibfnamefont {E.~P. A.~M.}\ \bibnamefont {Bakkers}}, \ and\ \bibinfo
  {author} {\bibfnamefont {L.~P.}\ \bibnamefont {Kouwenhoven}},\ }\href
  {\doibase 10.1126/science.1222360} {\bibfield  {journal} {\bibinfo  {journal}
  {Science}\ }\textbf {\bibinfo {volume} {{\bf 336}}},\ \bibinfo {pages} {1003}
  (\bibinfo {year} {2012})}\BibitemShut {NoStop}%
\bibitem [{\citenamefont {Deng}\ \emph {et~al.}(2012)\citenamefont {Deng},
  \citenamefont {Yu}, \citenamefont {Huang}, \citenamefont {Larsson},
  \citenamefont {Caroff},\ and\ \citenamefont {Xu}}]{Deng:2012}%
  \BibitemOpen
  \bibfield  {author} {\bibinfo {author} {\bibfnamefont {M.~T.}\ \bibnamefont
  {Deng}}, \bibinfo {author} {\bibfnamefont {C.~L.}\ \bibnamefont {Yu}},
  \bibinfo {author} {\bibfnamefont {G.~Y.}\ \bibnamefont {Huang}}, \bibinfo
  {author} {\bibfnamefont {M.}~\bibnamefont {Larsson}}, \bibinfo {author}
  {\bibfnamefont {P.}~\bibnamefont {Caroff}}, \ and\ \bibinfo {author}
  {\bibfnamefont {H.~Q.}\ \bibnamefont {Xu}},\ }\href {\doibase
  10.1021/nl303758w} {\bibfield  {journal} {\bibinfo  {journal} {Nano Letters}\
  }\textbf {\bibinfo {volume} {{\bf 12}}},\ \bibinfo {pages} {6414} (\bibinfo
  {year} {2012})}\BibitemShut {NoStop}%
\bibitem [{\citenamefont {Das}\ \emph {et~al.}(2012)\citenamefont {Das},
  \citenamefont {Ronen}, \citenamefont {Most}, \citenamefont {Oreg},
  \citenamefont {Heiblum},\ and\ \citenamefont {Shtrikman}}]{Das:2012}%
  \BibitemOpen
  \bibfield  {author} {\bibinfo {author} {\bibfnamefont {A.}~\bibnamefont
  {Das}}, \bibinfo {author} {\bibfnamefont {Y.}~\bibnamefont {Ronen}}, \bibinfo
  {author} {\bibfnamefont {Y.}~\bibnamefont {Most}}, \bibinfo {author}
  {\bibfnamefont {Y.}~\bibnamefont {Oreg}}, \bibinfo {author} {\bibfnamefont
  {M.}~\bibnamefont {Heiblum}}, \ and\ \bibinfo {author} {\bibfnamefont
  {H.}~\bibnamefont {Shtrikman}},\ }\href {http://dx.doi.org/10.1038/nphys2479}
  {\bibfield  {journal} {\bibinfo  {journal} {Nat. Phys.}\ }\textbf {\bibinfo
  {volume} {{\bf 8}}},\ \bibinfo {pages} {887} (\bibinfo {year}
  {2012})}\BibitemShut {NoStop}%
\bibitem [{\citenamefont {Churchill}\ \emph {et~al.}(2013)\citenamefont
  {Churchill}, \citenamefont {Fatemi}, \citenamefont {Grove-Rasmussen},
  \citenamefont {Deng}, \citenamefont {Caroff}, \citenamefont {Xu},\ and\
  \citenamefont {Marcus}}]{Churchill:2013}%
  \BibitemOpen
  \bibfield  {author} {\bibinfo {author} {\bibfnamefont {H.~O.~H.}\
  \bibnamefont {Churchill}}, \bibinfo {author} {\bibfnamefont {V.}~\bibnamefont
  {Fatemi}}, \bibinfo {author} {\bibfnamefont {K.}~\bibnamefont
  {Grove-Rasmussen}}, \bibinfo {author} {\bibfnamefont {M.~T.}\ \bibnamefont
  {Deng}}, \bibinfo {author} {\bibfnamefont {P.}~\bibnamefont {Caroff}},
  \bibinfo {author} {\bibfnamefont {H.~Q.}\ \bibnamefont {Xu}}, \ and\ \bibinfo
  {author} {\bibfnamefont {C.~M.}\ \bibnamefont {Marcus}},\ }\href
  {http://link.aps.org/doi/10.1103/PhysRevB.87.241401} {\bibfield  {journal}
  {\bibinfo  {journal} {Phys. Rev. B}\ }\textbf {\bibinfo {volume} {{\bf
  87}}},\ \bibinfo {pages} {241401} (\bibinfo {year} {2013})}\BibitemShut
  {NoStop}%
\bibitem [{\citenamefont {Finck}\ \emph {et~al.}(2013)\citenamefont {Finck},
  \citenamefont {Van~Harlingen}, \citenamefont {Mohseni}, \citenamefont
  {Jung},\ and\ \citenamefont {Li}}]{Finck:2013}%
  \BibitemOpen
  \bibfield  {author} {\bibinfo {author} {\bibfnamefont {A.~D.~K.}\
  \bibnamefont {Finck}}, \bibinfo {author} {\bibfnamefont {D.~J.}\ \bibnamefont
  {Van~Harlingen}}, \bibinfo {author} {\bibfnamefont {P.~K.}\ \bibnamefont
  {Mohseni}}, \bibinfo {author} {\bibfnamefont {K.}~\bibnamefont {Jung}}, \
  and\ \bibinfo {author} {\bibfnamefont {X.}~\bibnamefont {Li}},\ }\href
  {\doibase 10.1103/PhysRevLett.110.126406} {\bibfield  {journal} {\bibinfo
  {journal} {Phys. Rev. Lett.}\ }\textbf {\bibinfo {volume} {{\bf 110}}},\
  \bibinfo {pages} {126406} (\bibinfo {year} {2013})}\BibitemShut {NoStop}%
\bibitem [{\citenamefont {Lutchyn}\ \emph {et~al.}(2018)\citenamefont
  {Lutchyn}, \citenamefont {Bakkers}, \citenamefont {Kouwenhoven},
  \citenamefont {Krogstrup}, \citenamefont {Marcus},\ and\ \citenamefont
  {Oreg}}]{Lutchyn:2018}%
  \BibitemOpen
  \bibfield  {author} {\bibinfo {author} {\bibfnamefont {R.~M.}\ \bibnamefont
  {Lutchyn}}, \bibinfo {author} {\bibfnamefont {E.~P. A.~M.}\ \bibnamefont
  {Bakkers}}, \bibinfo {author} {\bibfnamefont {L.~P.}\ \bibnamefont
  {Kouwenhoven}}, \bibinfo {author} {\bibfnamefont {P.}~\bibnamefont
  {Krogstrup}}, \bibinfo {author} {\bibfnamefont {C.~M.}\ \bibnamefont
  {Marcus}}, \ and\ \bibinfo {author} {\bibfnamefont {Y.}~\bibnamefont
  {Oreg}},\ }\href {https://doi.org/10.1038/s41578-018-0003-1} {\bibfield
  {journal} {\bibinfo  {journal} {Nature Reviews Materials}\ }\textbf {\bibinfo
  {volume} {{\bf 3}}},\ \bibinfo {pages} {52} (\bibinfo {year}
  {2018})}\BibitemShut {NoStop}%
\bibitem [{\citenamefont {Chang}\ \emph {et~al.}(2015)\citenamefont {Chang},
  \citenamefont {Albrecht}, \citenamefont {Jespersen}, \citenamefont
  {Kuemmeth}, \citenamefont {Krogstrup}, \citenamefont {Nyg{\aa}rd},\ and\
  \citenamefont {Marcus}}]{Chang:2015}%
  \BibitemOpen
  \bibfield  {author} {\bibinfo {author} {\bibfnamefont {W.}~\bibnamefont
  {Chang}}, \bibinfo {author} {\bibfnamefont {S.~M.}\ \bibnamefont {Albrecht}},
  \bibinfo {author} {\bibfnamefont {T.~S.}\ \bibnamefont {Jespersen}}, \bibinfo
  {author} {\bibfnamefont {F.}~\bibnamefont {Kuemmeth}}, \bibinfo {author}
  {\bibfnamefont {P.}~\bibnamefont {Krogstrup}}, \bibinfo {author}
  {\bibfnamefont {J.}~\bibnamefont {Nyg{\aa}rd}}, \ and\ \bibinfo {author}
  {\bibfnamefont {C.~M.}\ \bibnamefont {Marcus}},\ }\href
  {http://dx.doi.org/10.1038/nnano.2014.306} {\bibfield  {journal} {\bibinfo
  {journal} {Nat. Nano.}\ }\textbf {\bibinfo {volume} {{\bf 10}}},\ \bibinfo
  {pages} {232} (\bibinfo {year} {2015})}\BibitemShut {NoStop}%
\bibitem [{\citenamefont {Gazibegovic}\ \emph {et~al.}(2017)\citenamefont
  {Gazibegovic}, \citenamefont {Car}, \citenamefont {Zhang}, \citenamefont
  {Balk}, \citenamefont {Logan}, \citenamefont {de~Moor}, \citenamefont
  {Cassidy}, \citenamefont {Schmits}, \citenamefont {Xu}, \citenamefont {Wang},
  \citenamefont {Krogstrup}, \citenamefont {Op~het Veld}, \citenamefont {Zuo},
  \citenamefont {Vos}, \citenamefont {Shen}, \citenamefont {Bouman},
  \citenamefont {Shojaei}, \citenamefont {Pennachio}, \citenamefont {Lee},
  \citenamefont {van Veldhoven}, \citenamefont {Koelling}, \citenamefont
  {Verheijen}, \citenamefont {Kouwenhoven}, \citenamefont {Palmstr{\o}m},\ and\
  \citenamefont {Bakkers}}]{Gazibegovic:2017}%
  \BibitemOpen
  \bibfield  {author} {\bibinfo {author} {\bibfnamefont {S.}~\bibnamefont
  {Gazibegovic}}, \bibinfo {author} {\bibfnamefont {D.}~\bibnamefont {Car}},
  \bibinfo {author} {\bibfnamefont {H.}~\bibnamefont {Zhang}}, \bibinfo
  {author} {\bibfnamefont {S.~C.}\ \bibnamefont {Balk}}, \bibinfo {author}
  {\bibfnamefont {J.~A.}\ \bibnamefont {Logan}}, \bibinfo {author}
  {\bibfnamefont {M.~W.~A.}\ \bibnamefont {de~Moor}}, \bibinfo {author}
  {\bibfnamefont {M.~C.}\ \bibnamefont {Cassidy}}, \bibinfo {author}
  {\bibfnamefont {R.}~\bibnamefont {Schmits}}, \bibinfo {author} {\bibfnamefont
  {D.}~\bibnamefont {Xu}}, \bibinfo {author} {\bibfnamefont {G.}~\bibnamefont
  {Wang}}, \bibinfo {author} {\bibfnamefont {P.}~\bibnamefont {Krogstrup}},
  \bibinfo {author} {\bibfnamefont {R.~L.~M.}\ \bibnamefont {Op~het Veld}},
  \bibinfo {author} {\bibfnamefont {K.}~\bibnamefont {Zuo}}, \bibinfo {author}
  {\bibfnamefont {Y.}~\bibnamefont {Vos}}, \bibinfo {author} {\bibfnamefont
  {J.}~\bibnamefont {Shen}}, \bibinfo {author} {\bibfnamefont {D.}~\bibnamefont
  {Bouman}}, \bibinfo {author} {\bibfnamefont {B.}~\bibnamefont {Shojaei}},
  \bibinfo {author} {\bibfnamefont {D.}~\bibnamefont {Pennachio}}, \bibinfo
  {author} {\bibfnamefont {J.~S.}\ \bibnamefont {Lee}}, \bibinfo {author}
  {\bibfnamefont {P.~J.}\ \bibnamefont {van Veldhoven}}, \bibinfo {author}
  {\bibfnamefont {S.}~\bibnamefont {Koelling}}, \bibinfo {author}
  {\bibfnamefont {M.~A.}\ \bibnamefont {Verheijen}}, \bibinfo {author}
  {\bibfnamefont {L.~P.}\ \bibnamefont {Kouwenhoven}}, \bibinfo {author}
  {\bibfnamefont {C.~J.}\ \bibnamefont {Palmstr{\o}m}}, \ and\ \bibinfo
  {author} {\bibfnamefont {E.~P. A.~M.}\ \bibnamefont {Bakkers}},\ }\href
  {http://dx.doi.org/10.1038/nature23468} {\bibfield  {journal} {\bibinfo
  {journal} {Nature}\ }\textbf {\bibinfo {volume} {{\bf 548}}},\ \bibinfo
  {pages} {434} (\bibinfo {year} {2017})}\BibitemShut {NoStop}%
\bibitem [{\citenamefont {Kjaergaard}\ \emph {et~al.}(2016)\citenamefont
  {Kjaergaard}, \citenamefont {Nichele}, \citenamefont {Suominen},
  \citenamefont {Nowak}, \citenamefont {Wimmer}, \citenamefont {Akhmerov},
  \citenamefont {Folk}, \citenamefont {Flensberg}, \citenamefont {Shabani},
  \citenamefont {Palmstr{\o}m},\ and\ \citenamefont
  {Marcus}}]{Kjaergaard:2016}%
  \BibitemOpen
  \bibfield  {author} {\bibinfo {author} {\bibfnamefont {M.}~\bibnamefont
  {Kjaergaard}}, \bibinfo {author} {\bibfnamefont {F.}~\bibnamefont {Nichele}},
  \bibinfo {author} {\bibfnamefont {H.~J.}\ \bibnamefont {Suominen}}, \bibinfo
  {author} {\bibfnamefont {M.~P.}\ \bibnamefont {Nowak}}, \bibinfo {author}
  {\bibfnamefont {M.}~\bibnamefont {Wimmer}}, \bibinfo {author} {\bibfnamefont
  {A.~R.}\ \bibnamefont {Akhmerov}}, \bibinfo {author} {\bibfnamefont {J.~A.}\
  \bibnamefont {Folk}}, \bibinfo {author} {\bibfnamefont {K.}~\bibnamefont
  {Flensberg}}, \bibinfo {author} {\bibfnamefont {J.}~\bibnamefont {Shabani}},
  \bibinfo {author} {\bibfnamefont {C.~J.}\ \bibnamefont {Palmstr{\o}m}}, \
  and\ \bibinfo {author} {\bibfnamefont {C.~M.}\ \bibnamefont {Marcus}},\
  }\href {http://dx.doi.org/10.1038/ncomms12841} {\bibfield  {journal}
  {\bibinfo  {journal} {Nat. Commun.}\ }\textbf {\bibinfo {volume} {{\bf 7}}},\
  \bibinfo {pages} {12841} (\bibinfo {year} {2016})}\BibitemShut {NoStop}%
\bibitem [{\citenamefont {Shabani}\ \emph {et~al.}(2016)\citenamefont
  {Shabani}, \citenamefont {Kjaergaard}, \citenamefont {Suominen},
  \citenamefont {Kim}, \citenamefont {Nichele}, \citenamefont {Pakrouski},
  \citenamefont {Stankevic}, \citenamefont {Lutchyn}, \citenamefont
  {Krogstrup}, \citenamefont {Feidenhans'l}, \citenamefont {Kraemer},
  \citenamefont {Nayak}, \citenamefont {Troyer}, \citenamefont {Marcus},\ and\
  \citenamefont {Palmstr\o{}m}}]{Shabani:2016}%
  \BibitemOpen
  \bibfield  {author} {\bibinfo {author} {\bibfnamefont {J.}~\bibnamefont
  {Shabani}}, \bibinfo {author} {\bibfnamefont {M.}~\bibnamefont {Kjaergaard}},
  \bibinfo {author} {\bibfnamefont {H.~J.}\ \bibnamefont {Suominen}}, \bibinfo
  {author} {\bibfnamefont {Y.}~\bibnamefont {Kim}}, \bibinfo {author}
  {\bibfnamefont {F.}~\bibnamefont {Nichele}}, \bibinfo {author} {\bibfnamefont
  {K.}~\bibnamefont {Pakrouski}}, \bibinfo {author} {\bibfnamefont
  {T.}~\bibnamefont {Stankevic}}, \bibinfo {author} {\bibfnamefont {R.~M.}\
  \bibnamefont {Lutchyn}}, \bibinfo {author} {\bibfnamefont {P.}~\bibnamefont
  {Krogstrup}}, \bibinfo {author} {\bibfnamefont {R.}~\bibnamefont
  {Feidenhans'l}}, \bibinfo {author} {\bibfnamefont {S.}~\bibnamefont
  {Kraemer}}, \bibinfo {author} {\bibfnamefont {C.}~\bibnamefont {Nayak}},
  \bibinfo {author} {\bibfnamefont {M.}~\bibnamefont {Troyer}}, \bibinfo
  {author} {\bibfnamefont {C.~M.}\ \bibnamefont {Marcus}}, \ and\ \bibinfo
  {author} {\bibfnamefont {C.~J.}\ \bibnamefont {Palmstr\o{}m}},\ }\href
  {https://link.aps.org/doi/10.1103/PhysRevB.93.155402} {\bibfield  {journal}
  {\bibinfo  {journal} {Phys. Rev. B}\ }\textbf {\bibinfo {volume} {{\bf
  93}}},\ \bibinfo {pages} {155402} (\bibinfo {year} {2016})}\BibitemShut
  {NoStop}%
\bibitem [{\citenamefont {Deng}\ \emph {et~al.}(2016)\citenamefont {Deng},
  \citenamefont {Vaitiekenas}, \citenamefont {Hansen}, \citenamefont {Danon},
  \citenamefont {Leijnse}, \citenamefont {Flensberg}, \citenamefont
  {Nyg{\aa}rd}, \citenamefont {Krogstrup},\ and\ \citenamefont
  {Marcus}}]{Deng:2016}%
  \BibitemOpen
  \bibfield  {author} {\bibinfo {author} {\bibfnamefont {M.~T.}\ \bibnamefont
  {Deng}}, \bibinfo {author} {\bibfnamefont {S.}~\bibnamefont {Vaitiekenas}},
  \bibinfo {author} {\bibfnamefont {E.~B.}\ \bibnamefont {Hansen}}, \bibinfo
  {author} {\bibfnamefont {J.}~\bibnamefont {Danon}}, \bibinfo {author}
  {\bibfnamefont {M.}~\bibnamefont {Leijnse}}, \bibinfo {author} {\bibfnamefont
  {K.}~\bibnamefont {Flensberg}}, \bibinfo {author} {\bibfnamefont
  {J.}~\bibnamefont {Nyg{\aa}rd}}, \bibinfo {author} {\bibfnamefont
  {P.}~\bibnamefont {Krogstrup}}, \ and\ \bibinfo {author} {\bibfnamefont
  {C.~M.}\ \bibnamefont {Marcus}},\ }\href
  {http://science.sciencemag.org/content/354/6319/1557} {\bibfield  {journal}
  {\bibinfo  {journal} {Science}\ }\textbf {\bibinfo {volume} {{\bf 354}}},\
  \bibinfo {pages} {1557} (\bibinfo {year} {2016})}\BibitemShut {NoStop}%
\bibitem [{\citenamefont {Zhang}\ \emph {et~al.}(2018)\citenamefont {Zhang},
  \citenamefont {Liu}, \citenamefont {Gazibegovic}, \citenamefont {Xu},
  \citenamefont {Logan}, \citenamefont {Wang}, \citenamefont {van Loo},
  \citenamefont {Bommer}, \citenamefont {de~Moor}, \citenamefont {Car},
  \citenamefont {Op~het Veld}, \citenamefont {van Veldhoven}, \citenamefont
  {Koelling}, \citenamefont {Verheijen}, \citenamefont {Pendharkar},
  \citenamefont {Pennachio}, \citenamefont {Shojaei}, \citenamefont {Lee},
  \citenamefont {Palmstr{\o}m}, \citenamefont {Bakkers}, \citenamefont
  {Sarma},\ and\ \citenamefont {Kouwenhoven}}]{Zhang:2018}%
  \BibitemOpen
  \bibfield  {author} {\bibinfo {author} {\bibfnamefont {H.}~\bibnamefont
  {Zhang}}, \bibinfo {author} {\bibfnamefont {C.-X.}\ \bibnamefont {Liu}},
  \bibinfo {author} {\bibfnamefont {S.}~\bibnamefont {Gazibegovic}}, \bibinfo
  {author} {\bibfnamefont {D.}~\bibnamefont {Xu}}, \bibinfo {author}
  {\bibfnamefont {J.~A.}\ \bibnamefont {Logan}}, \bibinfo {author}
  {\bibfnamefont {G.}~\bibnamefont {Wang}}, \bibinfo {author} {\bibfnamefont
  {N.}~\bibnamefont {van Loo}}, \bibinfo {author} {\bibfnamefont {J.~D.~S.}\
  \bibnamefont {Bommer}}, \bibinfo {author} {\bibfnamefont {M.~W.~A.}\
  \bibnamefont {de~Moor}}, \bibinfo {author} {\bibfnamefont {D.}~\bibnamefont
  {Car}}, \bibinfo {author} {\bibfnamefont {R.~L.~M.}\ \bibnamefont {Op~het
  Veld}}, \bibinfo {author} {\bibfnamefont {P.~J.}\ \bibnamefont {van
  Veldhoven}}, \bibinfo {author} {\bibfnamefont {S.}~\bibnamefont {Koelling}},
  \bibinfo {author} {\bibfnamefont {M.~A.}\ \bibnamefont {Verheijen}}, \bibinfo
  {author} {\bibfnamefont {M.}~\bibnamefont {Pendharkar}}, \bibinfo {author}
  {\bibfnamefont {D.~J.}\ \bibnamefont {Pennachio}}, \bibinfo {author}
  {\bibfnamefont {B.}~\bibnamefont {Shojaei}}, \bibinfo {author} {\bibfnamefont
  {J.~S.}\ \bibnamefont {Lee}}, \bibinfo {author} {\bibfnamefont {C.~J.}\
  \bibnamefont {Palmstr{\o}m}}, \bibinfo {author} {\bibfnamefont {E.~P. A.~M.}\
  \bibnamefont {Bakkers}}, \bibinfo {author} {\bibfnamefont {S.~D.}\
  \bibnamefont {Sarma}}, \ and\ \bibinfo {author} {\bibfnamefont {L.~P.}\
  \bibnamefont {Kouwenhoven}},\ }\href
  {https://www.nature.com/articles/nature26142} {\bibfield  {journal} {\bibinfo
   {journal} {Nature}\ }\textbf {\bibinfo {volume} {{\bf 556}}},\ \bibinfo
  {pages} {74 EP } (\bibinfo {year} {2018})}\BibitemShut {NoStop}%
\bibitem [{\citenamefont {Vaitiek\ifmmode~\dot{e}\else \.{e}\fi{}nas}\ \emph
  {et~al.}(2018)\citenamefont {Vaitiek\ifmmode~\dot{e}\else \.{e}\fi{}nas},
  \citenamefont {Deng}, \citenamefont {Nyg\aa{}rd}, \citenamefont {Krogstrup},\
  and\ \citenamefont {Marcus}}]{Vaitiekenas:2018}%
  \BibitemOpen
  \bibfield  {author} {\bibinfo {author} {\bibfnamefont {S.}~\bibnamefont
  {Vaitiek\ifmmode~\dot{e}\else \.{e}\fi{}nas}}, \bibinfo {author}
  {\bibfnamefont {M.-T.}\ \bibnamefont {Deng}}, \bibinfo {author}
  {\bibfnamefont {J.}~\bibnamefont {Nyg\aa{}rd}}, \bibinfo {author}
  {\bibfnamefont {P.}~\bibnamefont {Krogstrup}}, \ and\ \bibinfo {author}
  {\bibfnamefont {C.~M.}\ \bibnamefont {Marcus}},\ }\href
  {https://link.aps.org/doi/10.1103/PhysRevLett.121.037703} {\bibfield
  {journal} {\bibinfo  {journal} {Phys. Rev. Lett.}\ }\textbf {\bibinfo
  {volume} {{\bf 121}}},\ \bibinfo {pages} {037703} (\bibinfo {year}
  {2018})}\BibitemShut {NoStop}%
\bibitem [{\citenamefont {Deng}\ \emph {et~al.}(2018)\citenamefont {Deng},
  \citenamefont {Vaitiek\ifmmode~\dot{e}\else \.{e}\fi{}nas}, \citenamefont
  {Prada}, \citenamefont {San-Jose}, \citenamefont {Nyg\aa{}rd}, \citenamefont
  {Krogstrup}, \citenamefont {Aguado},\ and\ \citenamefont
  {Marcus}}]{Deng:2018}%
  \BibitemOpen
  \bibfield  {author} {\bibinfo {author} {\bibfnamefont {M.-T.}\ \bibnamefont
  {Deng}}, \bibinfo {author} {\bibfnamefont {S.}~\bibnamefont
  {Vaitiek\ifmmode~\dot{e}\else \.{e}\fi{}nas}}, \bibinfo {author}
  {\bibfnamefont {E.}~\bibnamefont {Prada}}, \bibinfo {author} {\bibfnamefont
  {P.}~\bibnamefont {San-Jose}}, \bibinfo {author} {\bibfnamefont
  {J.}~\bibnamefont {Nyg\aa{}rd}}, \bibinfo {author} {\bibfnamefont
  {P.}~\bibnamefont {Krogstrup}}, \bibinfo {author} {\bibfnamefont
  {R.}~\bibnamefont {Aguado}}, \ and\ \bibinfo {author} {\bibfnamefont {C.~M.}\
  \bibnamefont {Marcus}},\ }\href
  {https://link.aps.org/doi/10.1103/PhysRevB.98.085125} {\bibfield  {journal}
  {\bibinfo  {journal} {Phys. Rev. B}\ }\textbf {\bibinfo {volume} {{\bf
  98}}},\ \bibinfo {pages} {085125} (\bibinfo {year} {2018})}\BibitemShut
  {NoStop}%
\bibitem [{\citenamefont {de~Moor}\ \emph {et~al.}(2018)\citenamefont
  {de~Moor}, \citenamefont {Bommer}, \citenamefont {Xu}, \citenamefont
  {Winkler}, \citenamefont {Antipov}, \citenamefont {Bargerbos}, \citenamefont
  {Wang}, \citenamefont {van Loo}, \citenamefont {het Veld}, \citenamefont
  {Gazibegovic}, \citenamefont {Car}, \citenamefont {Logan}, \citenamefont
  {Pendharkar}, \citenamefont {Lee}, \citenamefont {Bakkers}, \citenamefont
  {Palmstr{\o}m}, \citenamefont {Lutchyn}, \citenamefont {Kouwenhoven},\ and\
  \citenamefont {Zhang}}]{deMoor:2018}%
  \BibitemOpen
  \bibfield  {author} {\bibinfo {author} {\bibfnamefont {M.~W.~A.}\
  \bibnamefont {de~Moor}}, \bibinfo {author} {\bibfnamefont {J.~D.~S.}\
  \bibnamefont {Bommer}}, \bibinfo {author} {\bibfnamefont {D.}~\bibnamefont
  {Xu}}, \bibinfo {author} {\bibfnamefont {G.~W.}\ \bibnamefont {Winkler}},
  \bibinfo {author} {\bibfnamefont {A.~E.}\ \bibnamefont {Antipov}}, \bibinfo
  {author} {\bibfnamefont {A.}~\bibnamefont {Bargerbos}}, \bibinfo {author}
  {\bibfnamefont {G.}~\bibnamefont {Wang}}, \bibinfo {author} {\bibfnamefont
  {N.}~\bibnamefont {van Loo}}, \bibinfo {author} {\bibfnamefont {R.~L. M.~O.}\
  \bibnamefont {het Veld}}, \bibinfo {author} {\bibfnamefont {S.}~\bibnamefont
  {Gazibegovic}}, \bibinfo {author} {\bibfnamefont {D.}~\bibnamefont {Car}},
  \bibinfo {author} {\bibfnamefont {J.~A.}\ \bibnamefont {Logan}}, \bibinfo
  {author} {\bibfnamefont {M.}~\bibnamefont {Pendharkar}}, \bibinfo {author}
  {\bibfnamefont {J.~S.}\ \bibnamefont {Lee}}, \bibinfo {author} {\bibfnamefont
  {E.~P. A.~M.}\ \bibnamefont {Bakkers}}, \bibinfo {author} {\bibfnamefont
  {C.~J.}\ \bibnamefont {Palmstr{\o}m}}, \bibinfo {author} {\bibfnamefont
  {R.~M.}\ \bibnamefont {Lutchyn}}, \bibinfo {author} {\bibfnamefont {L.~P.}\
  \bibnamefont {Kouwenhoven}}, \ and\ \bibinfo {author} {\bibfnamefont
  {H.}~\bibnamefont {Zhang}},\ }\href
  {http://stacks.iop.org/1367-2630/20/i=10/a=103049} {\bibfield  {journal}
  {\bibinfo  {journal} {New J. Phys.}\ }\textbf {\bibinfo {volume} {{\bf
  20}}},\ \bibinfo {pages} {103049} (\bibinfo {year} {2018})}\BibitemShut
  {NoStop}%
\bibitem [{\citenamefont {Suominen}\ \emph {et~al.}(2017)\citenamefont
  {Suominen}, \citenamefont {Kjaergaard}, \citenamefont {Hamilton},
  \citenamefont {Shabani}, \citenamefont {Palmstr\o{}m}, \citenamefont
  {Marcus},\ and\ \citenamefont {Nichele}}]{Suominen:2017}%
  \BibitemOpen
  \bibfield  {author} {\bibinfo {author} {\bibfnamefont {H.~J.}\ \bibnamefont
  {Suominen}}, \bibinfo {author} {\bibfnamefont {M.}~\bibnamefont
  {Kjaergaard}}, \bibinfo {author} {\bibfnamefont {A.~R.}\ \bibnamefont
  {Hamilton}}, \bibinfo {author} {\bibfnamefont {J.}~\bibnamefont {Shabani}},
  \bibinfo {author} {\bibfnamefont {C.~J.}\ \bibnamefont {Palmstr\o{}m}},
  \bibinfo {author} {\bibfnamefont {C.~M.}\ \bibnamefont {Marcus}}, \ and\
  \bibinfo {author} {\bibfnamefont {F.}~\bibnamefont {Nichele}},\ }\href
  {https://link.aps.org/doi/10.1103/PhysRevLett.119.176805} {\bibfield
  {journal} {\bibinfo  {journal} {Phys. Rev. Lett.}\ }\textbf {\bibinfo
  {volume} {{\bf 119}}},\ \bibinfo {pages} {176805} (\bibinfo {year}
  {2017})}\BibitemShut {NoStop}%
\bibitem [{\citenamefont {Nichele}\ \emph {et~al.}(2017)\citenamefont
  {Nichele}, \citenamefont {Drachmann}, \citenamefont {Whiticar}, \citenamefont
  {O'Farrell}, \citenamefont {Suominen}, \citenamefont {Fornieri},
  \citenamefont {Wang}, \citenamefont {Gardner}, \citenamefont {Thomas},
  \citenamefont {Hatke}, \citenamefont {Krogstrup}, \citenamefont {Manfra},
  \citenamefont {Flensberg},\ and\ \citenamefont {Marcus}}]{Nichele:2017}%
  \BibitemOpen
  \bibfield  {author} {\bibinfo {author} {\bibfnamefont {F.}~\bibnamefont
  {Nichele}}, \bibinfo {author} {\bibfnamefont {A.~C.~C.}\ \bibnamefont
  {Drachmann}}, \bibinfo {author} {\bibfnamefont {A.~M.}\ \bibnamefont
  {Whiticar}}, \bibinfo {author} {\bibfnamefont {E.~C.~T.}\ \bibnamefont
  {O'Farrell}}, \bibinfo {author} {\bibfnamefont {H.~J.}\ \bibnamefont
  {Suominen}}, \bibinfo {author} {\bibfnamefont {A.}~\bibnamefont {Fornieri}},
  \bibinfo {author} {\bibfnamefont {T.}~\bibnamefont {Wang}}, \bibinfo {author}
  {\bibfnamefont {G.~C.}\ \bibnamefont {Gardner}}, \bibinfo {author}
  {\bibfnamefont {C.}~\bibnamefont {Thomas}}, \bibinfo {author} {\bibfnamefont
  {A.~T.}\ \bibnamefont {Hatke}}, \bibinfo {author} {\bibfnamefont
  {P.}~\bibnamefont {Krogstrup}}, \bibinfo {author} {\bibfnamefont {M.~J.}\
  \bibnamefont {Manfra}}, \bibinfo {author} {\bibfnamefont {K.}~\bibnamefont
  {Flensberg}}, \ and\ \bibinfo {author} {\bibfnamefont {C.~M.}\ \bibnamefont
  {Marcus}},\ }\href {https://link.aps.org/doi/10.1103/PhysRevLett.119.136803}
  {\bibfield  {journal} {\bibinfo  {journal} {Phys. Rev. Lett.}\ }\textbf
  {\bibinfo {volume} {{\bf 119}}},\ \bibinfo {pages} {136803} (\bibinfo {year}
  {2017})}\BibitemShut {NoStop}%
\bibitem [{\citenamefont {Reeg}\ \emph {et~al.}(2017)\citenamefont {Reeg},
  \citenamefont {Loss},\ and\ \citenamefont {Klinovaja}}]{Reeg:2017_3}%
  \BibitemOpen
  \bibfield  {author} {\bibinfo {author} {\bibfnamefont {C.}~\bibnamefont
  {Reeg}}, \bibinfo {author} {\bibfnamefont {D.}~\bibnamefont {Loss}}, \ and\
  \bibinfo {author} {\bibfnamefont {J.}~\bibnamefont {Klinovaja}},\ }\href
  {https://link.aps.org/doi/10.1103/PhysRevB.96.125426} {\bibfield  {journal}
  {\bibinfo  {journal} {Phys. Rev. B}\ }\textbf {\bibinfo {volume} {{\bf
  96}}},\ \bibinfo {pages} {125426} (\bibinfo {year} {2017})}\BibitemShut
  {NoStop}%
\bibitem [{\citenamefont {Reeg}\ \emph
  {et~al.}(2018{\natexlab{a}})\citenamefont {Reeg}, \citenamefont {Loss},\ and\
  \citenamefont {Klinovaja}}]{Reeg:2018}%
  \BibitemOpen
  \bibfield  {author} {\bibinfo {author} {\bibfnamefont {C.}~\bibnamefont
  {Reeg}}, \bibinfo {author} {\bibfnamefont {D.}~\bibnamefont {Loss}}, \ and\
  \bibinfo {author} {\bibfnamefont {J.}~\bibnamefont {Klinovaja}},\ }\href
  {https://link.aps.org/doi/10.1103/PhysRevB.97.165425} {\bibfield  {journal}
  {\bibinfo  {journal} {Phys. Rev. B}\ }\textbf {\bibinfo {volume} {{\bf
  97}}},\ \bibinfo {pages} {165425} (\bibinfo {year}
  {2018}{\natexlab{a}})}\BibitemShut {NoStop}%
\bibitem [{\citenamefont {Reeg}\ \emph
  {et~al.}(2018{\natexlab{b}})\citenamefont {Reeg}, \citenamefont {Loss},\ and\
  \citenamefont {Klinovaja}}]{Reeg:2018_2}%
  \BibitemOpen
  \bibfield  {author} {\bibinfo {author} {\bibfnamefont {C.}~\bibnamefont
  {Reeg}}, \bibinfo {author} {\bibfnamefont {D.}~\bibnamefont {Loss}}, \ and\
  \bibinfo {author} {\bibfnamefont {J.}~\bibnamefont {Klinovaja}},\ }\href
  {https://www.beilstein-journals.org/bjnano/articles/9/118} {\bibfield
  {journal} {\bibinfo  {journal} {Beilstein J. Nanotechnol.}\ }\textbf
  {\bibinfo {volume} {{\bf 9}}},\ \bibinfo {pages} {1263} (\bibinfo {year}
  {2018}{\natexlab{b}})}\BibitemShut {NoStop}%
\bibitem [{\citenamefont {Sau}\ \emph {et~al.}(2010)\citenamefont {Sau},
  \citenamefont {Lutchyn}, \citenamefont {Tewari},\ and\ \citenamefont
  {Das~Sarma}}]{Sau:2010prox}%
  \BibitemOpen
  \bibfield  {author} {\bibinfo {author} {\bibfnamefont {J.~D.}\ \bibnamefont
  {Sau}}, \bibinfo {author} {\bibfnamefont {R.~M.}\ \bibnamefont {Lutchyn}},
  \bibinfo {author} {\bibfnamefont {S.}~\bibnamefont {Tewari}}, \ and\ \bibinfo
  {author} {\bibfnamefont {S.}~\bibnamefont {Das~Sarma}},\ }\href
  {http://link.aps.org/doi/10.1103/PhysRevB.82.094522} {\bibfield  {journal}
  {\bibinfo  {journal} {Phys. Rev. B}\ }\textbf {\bibinfo {volume} {{\bf
  82}}},\ \bibinfo {pages} {094522} (\bibinfo {year} {2010})}\BibitemShut
  {NoStop}%
\bibitem [{\citenamefont {Stanescu}\ \emph {et~al.}(2010)\citenamefont
  {Stanescu}, \citenamefont {Sau}, \citenamefont {Lutchyn},\ and\ \citenamefont
  {Das~Sarma}}]{Stanescu:2010}%
  \BibitemOpen
  \bibfield  {author} {\bibinfo {author} {\bibfnamefont {T.~D.}\ \bibnamefont
  {Stanescu}}, \bibinfo {author} {\bibfnamefont {J.~D.}\ \bibnamefont {Sau}},
  \bibinfo {author} {\bibfnamefont {R.~M.}\ \bibnamefont {Lutchyn}}, \ and\
  \bibinfo {author} {\bibfnamefont {S.}~\bibnamefont {Das~Sarma}},\ }\href
  {https://link.aps.org/doi/10.1103/PhysRevB.81.241310} {\bibfield  {journal}
  {\bibinfo  {journal} {Phys. Rev. B}\ }\textbf {\bibinfo {volume} {{\bf
  81}}},\ \bibinfo {pages} {241310} (\bibinfo {year} {2010})}\BibitemShut
  {NoStop}%
\bibitem [{\citenamefont {Potter}\ and\ \citenamefont
  {Lee}(2011)}]{Potter:2011}%
  \BibitemOpen
  \bibfield  {author} {\bibinfo {author} {\bibfnamefont {A.~C.}\ \bibnamefont
  {Potter}}\ and\ \bibinfo {author} {\bibfnamefont {P.~A.}\ \bibnamefont
  {Lee}},\ }\href {\doibase 10.1103/PhysRevB.83.184520} {\bibfield  {journal}
  {\bibinfo  {journal} {Phys. Rev. B}\ }\textbf {\bibinfo {volume} {{\bf
  83}}},\ \bibinfo {pages} {184520} (\bibinfo {year} {2011})}\BibitemShut
  {NoStop}%
\bibitem [{\citenamefont {Tkachov}(2013)}]{Tkachov:2013}%
  \BibitemOpen
  \bibfield  {author} {\bibinfo {author} {\bibfnamefont {G.}~\bibnamefont
  {Tkachov}},\ }\href {https://link.aps.org/doi/10.1103/PhysRevB.87.245422}
  {\bibfield  {journal} {\bibinfo  {journal} {Phys. Rev. B}\ }\textbf {\bibinfo
  {volume} {{\bf 87}}},\ \bibinfo {pages} {245422} (\bibinfo {year}
  {2013})}\BibitemShut {NoStop}%
\bibitem [{\citenamefont {Zyuzin}\ \emph {et~al.}(2013)\citenamefont {Zyuzin},
  \citenamefont {Rainis}, \citenamefont {Klinovaja},\ and\ \citenamefont
  {Loss}}]{Zyuzin:2013}%
  \BibitemOpen
  \bibfield  {author} {\bibinfo {author} {\bibfnamefont {A.~A.}\ \bibnamefont
  {Zyuzin}}, \bibinfo {author} {\bibfnamefont {D.}~\bibnamefont {Rainis}},
  \bibinfo {author} {\bibfnamefont {J.}~\bibnamefont {Klinovaja}}, \ and\
  \bibinfo {author} {\bibfnamefont {D.}~\bibnamefont {Loss}},\ }\href
  {http://link.aps.org/doi/10.1103/PhysRevLett.111.056802} {\bibfield
  {journal} {\bibinfo  {journal} {Phys. Rev. Lett.}\ }\textbf {\bibinfo
  {volume} {{\bf 111}}},\ \bibinfo {pages} {056802} (\bibinfo {year}
  {2013})}\BibitemShut {NoStop}%
\bibitem [{\citenamefont {Cole}\ \emph {et~al.}(2015)\citenamefont {Cole},
  \citenamefont {Das~Sarma},\ and\ \citenamefont {Stanescu}}]{Cole:2015}%
  \BibitemOpen
  \bibfield  {author} {\bibinfo {author} {\bibfnamefont {W.~S.}\ \bibnamefont
  {Cole}}, \bibinfo {author} {\bibfnamefont {S.}~\bibnamefont {Das~Sarma}}, \
  and\ \bibinfo {author} {\bibfnamefont {T.~D.}\ \bibnamefont {Stanescu}},\
  }\href {https://link.aps.org/doi/10.1103/PhysRevB.92.174511} {\bibfield
  {journal} {\bibinfo  {journal} {Phys. Rev. B}\ }\textbf {\bibinfo {volume}
  {{\bf 92}}},\ \bibinfo {pages} {174511} (\bibinfo {year} {2015})}\BibitemShut
  {NoStop}%
\bibitem [{\citenamefont {van Heck}\ \emph {et~al.}(2016)\citenamefont {van
  Heck}, \citenamefont {Lutchyn},\ and\ \citenamefont
  {Glazman}}]{vanHeck:2016}%
  \BibitemOpen
  \bibfield  {author} {\bibinfo {author} {\bibfnamefont {B.}~\bibnamefont {van
  Heck}}, \bibinfo {author} {\bibfnamefont {R.~M.}\ \bibnamefont {Lutchyn}}, \
  and\ \bibinfo {author} {\bibfnamefont {L.~I.}\ \bibnamefont {Glazman}},\
  }\href {http://link.aps.org/doi/10.1103/PhysRevB.93.235431} {\bibfield
  {journal} {\bibinfo  {journal} {Phys. Rev. B}\ }\textbf {\bibinfo {volume}
  {{\bf 93}}},\ \bibinfo {pages} {235431} (\bibinfo {year} {2016})}\BibitemShut
  {NoStop}%
\bibitem [{\citenamefont {Hell}\ \emph {et~al.}(2017)\citenamefont {Hell},
  \citenamefont {Flensberg},\ and\ \citenamefont {Leijnse}}]{Hell:2017}%
  \BibitemOpen
  \bibfield  {author} {\bibinfo {author} {\bibfnamefont {M.}~\bibnamefont
  {Hell}}, \bibinfo {author} {\bibfnamefont {K.}~\bibnamefont {Flensberg}}, \
  and\ \bibinfo {author} {\bibfnamefont {M.}~\bibnamefont {Leijnse}},\ }\href
  {https://link.aps.org/doi/10.1103/PhysRevB.96.035444} {\bibfield  {journal}
  {\bibinfo  {journal} {Phys. Rev. B}\ }\textbf {\bibinfo {volume} {{\bf
  96}}},\ \bibinfo {pages} {035444} (\bibinfo {year} {2017})}\BibitemShut
  {NoStop}%
\bibitem [{\citenamefont {Stanescu}\ and\ \citenamefont
  {Das~Sarma}(2017)}]{Stanescu:2017}%
  \BibitemOpen
  \bibfield  {author} {\bibinfo {author} {\bibfnamefont {T.~D.}\ \bibnamefont
  {Stanescu}}\ and\ \bibinfo {author} {\bibfnamefont {S.}~\bibnamefont
  {Das~Sarma}},\ }\href {https://link.aps.org/doi/10.1103/PhysRevB.96.014510}
  {\bibfield  {journal} {\bibinfo  {journal} {Phys. Rev. B}\ }\textbf {\bibinfo
  {volume} {{\bf 96}}},\ \bibinfo {pages} {014510} (\bibinfo {year}
  {2017})}\BibitemShut {NoStop}%
\bibitem [{\citenamefont {Reeg}\ and\ \citenamefont
  {Maslov}(2017)}]{Reeg:2017_2}%
  \BibitemOpen
  \bibfield  {author} {\bibinfo {author} {\bibfnamefont {C.}~\bibnamefont
  {Reeg}}\ and\ \bibinfo {author} {\bibfnamefont {D.~L.}\ \bibnamefont
  {Maslov}},\ }\href {https://link.aps.org/doi/10.1103/PhysRevB.95.205439}
  {\bibfield  {journal} {\bibinfo  {journal} {Phys. Rev. B}\ }\textbf {\bibinfo
  {volume} {{\bf 95}}},\ \bibinfo {pages} {205439} (\bibinfo {year}
  {2017})}\BibitemShut {NoStop}%
\bibitem [{\citenamefont {{Antipov}}\ \emph {et~al.}()\citenamefont
  {{Antipov}}, \citenamefont {{Bargerbos}}, \citenamefont {{Winkler}},
  \citenamefont {{Bauer}}, \citenamefont {{Rossi}},\ and\ \citenamefont
  {{Lutchyn}}}]{Antipov:2018}%
  \BibitemOpen
  \bibfield  {author} {\bibinfo {author} {\bibfnamefont {A.~E.}\ \bibnamefont
  {{Antipov}}}, \bibinfo {author} {\bibfnamefont {A.}~\bibnamefont
  {{Bargerbos}}}, \bibinfo {author} {\bibfnamefont {G.~W.}\ \bibnamefont
  {{Winkler}}}, \bibinfo {author} {\bibfnamefont {B.}~\bibnamefont {{Bauer}}},
  \bibinfo {author} {\bibfnamefont {E.}~\bibnamefont {{Rossi}}}, \ and\
  \bibinfo {author} {\bibfnamefont {R.~M.}\ \bibnamefont {{Lutchyn}}},\ }\href
  {https://arxiv.org/abs/1801.02616} {\ }\Eprint
  {http://arxiv.org/abs/1801.02616} {arXiv:1801.02616} \BibitemShut {NoStop}%
\bibitem [{\citenamefont {Woods}\ \emph {et~al.}(2018)\citenamefont {Woods},
  \citenamefont {Stanescu},\ and\ \citenamefont {Das~Sarma}}]{Woods:2018}%
  \BibitemOpen
  \bibfield  {author} {\bibinfo {author} {\bibfnamefont {B.~D.}\ \bibnamefont
  {Woods}}, \bibinfo {author} {\bibfnamefont {T.~D.}\ \bibnamefont {Stanescu}},
  \ and\ \bibinfo {author} {\bibfnamefont {S.}~\bibnamefont {Das~Sarma}},\
  }\href {https://link.aps.org/doi/10.1103/PhysRevB.98.035428} {\bibfield
  {journal} {\bibinfo  {journal} {Phys. Rev. B}\ }\textbf {\bibinfo {volume}
  {{\bf 98}}},\ \bibinfo {pages} {035428} (\bibinfo {year} {2018})}\BibitemShut
  {NoStop}%
\bibitem [{\citenamefont {{Mikkelsen}}\ \emph {et~al.}()\citenamefont
  {{Mikkelsen}}, \citenamefont {{Kotetes}}, \citenamefont {{Krogstrup}},\ and\
  \citenamefont {{Flensberg}}}]{Mikkelsen:2018}%
  \BibitemOpen
  \bibfield  {author} {\bibinfo {author} {\bibfnamefont {A.~E.~G.}\
  \bibnamefont {{Mikkelsen}}}, \bibinfo {author} {\bibfnamefont
  {P.}~\bibnamefont {{Kotetes}}}, \bibinfo {author} {\bibfnamefont
  {P.}~\bibnamefont {{Krogstrup}}}, \ and\ \bibinfo {author} {\bibfnamefont
  {K.}~\bibnamefont {{Flensberg}}},\ }\href {https://arxiv.org/abs/1801.03439}
  {\ }\Eprint {http://arxiv.org/abs/1801.03439} {arXiv:1801.03439} \BibitemShut
  {NoStop}%
\bibitem [{\citenamefont {Liu}\ \emph {et~al.}(2017)\citenamefont {Liu},
  \citenamefont {Sau}, \citenamefont {Stanescu},\ and\ \citenamefont
  {Das~Sarma}}]{Liu:2017}%
  \BibitemOpen
  \bibfield  {author} {\bibinfo {author} {\bibfnamefont {C.-X.}\ \bibnamefont
  {Liu}}, \bibinfo {author} {\bibfnamefont {J.~D.}\ \bibnamefont {Sau}},
  \bibinfo {author} {\bibfnamefont {T.~D.}\ \bibnamefont {Stanescu}}, \ and\
  \bibinfo {author} {\bibfnamefont {S.}~\bibnamefont {Das~Sarma}},\ }\href
  {https://link.aps.org/doi/10.1103/PhysRevB.96.075161} {\bibfield  {journal}
  {\bibinfo  {journal} {Phys. Rev. B}\ }\textbf {\bibinfo {volume} {{\bf
  96}}},\ \bibinfo {pages} {075161} (\bibinfo {year} {2017})}\BibitemShut
  {NoStop}%
\bibitem [{\citenamefont {Ptok}\ \emph {et~al.}(2017)\citenamefont {Ptok},
  \citenamefont {Kobia\l{}ka},\ and\ \citenamefont {Doma\ifmmode~\acute{n}\else
  \'{n}\fi{}ski}}]{Ptok:2017}%
  \BibitemOpen
  \bibfield  {author} {\bibinfo {author} {\bibfnamefont {A.}~\bibnamefont
  {Ptok}}, \bibinfo {author} {\bibfnamefont {A.}~\bibnamefont {Kobia\l{}ka}}, \
  and\ \bibinfo {author} {\bibfnamefont {T.}~\bibnamefont
  {Doma\ifmmode~\acute{n}\else \'{n}\fi{}ski}},\ }\href
  {https://link.aps.org/doi/10.1103/PhysRevB.96.195430} {\bibfield  {journal}
  {\bibinfo  {journal} {Phys. Rev. B}\ }\textbf {\bibinfo {volume} {{\bf
  96}}},\ \bibinfo {pages} {195430} (\bibinfo {year} {2017})}\BibitemShut
  {NoStop}%
\bibitem [{\citenamefont {Setiawan}\ \emph {et~al.}(2017)\citenamefont
  {Setiawan}, \citenamefont {Liu}, \citenamefont {Sau},\ and\ \citenamefont
  {Das~Sarma}}]{Setiawan:2017}%
  \BibitemOpen
  \bibfield  {author} {\bibinfo {author} {\bibfnamefont {F.}~\bibnamefont
  {Setiawan}}, \bibinfo {author} {\bibfnamefont {C.-X.}\ \bibnamefont {Liu}},
  \bibinfo {author} {\bibfnamefont {J.~D.}\ \bibnamefont {Sau}}, \ and\
  \bibinfo {author} {\bibfnamefont {S.}~\bibnamefont {Das~Sarma}},\ }\href
  {https://link.aps.org/doi/10.1103/PhysRevB.96.184520} {\bibfield  {journal}
  {\bibinfo  {journal} {Phys. Rev. B}\ }\textbf {\bibinfo {volume} {{\bf
  96}}},\ \bibinfo {pages} {184520} (\bibinfo {year} {2017})}\BibitemShut
  {NoStop}%
\bibitem [{\citenamefont {Moore}\ \emph {et~al.}(2018)\citenamefont {Moore},
  \citenamefont {Stanescu},\ and\ \citenamefont {Tewari}}]{Moore:2018}%
  \BibitemOpen
  \bibfield  {author} {\bibinfo {author} {\bibfnamefont {C.}~\bibnamefont
  {Moore}}, \bibinfo {author} {\bibfnamefont {T.~D.}\ \bibnamefont {Stanescu}},
  \ and\ \bibinfo {author} {\bibfnamefont {S.}~\bibnamefont {Tewari}},\ }\href
  {https://link.aps.org/doi/10.1103/PhysRevB.97.165302} {\bibfield  {journal}
  {\bibinfo  {journal} {Phys. Rev. B}\ }\textbf {\bibinfo {volume} {{\bf
  97}}},\ \bibinfo {pages} {165302} (\bibinfo {year} {2018})}\BibitemShut
  {NoStop}%
\bibitem [{\citenamefont {{Vuik}}\ \emph {et~al.}()\citenamefont {{Vuik}},
  \citenamefont {{Nijholt}}, \citenamefont {{Akhmerov}},\ and\ \citenamefont
  {{Wimmer}}}]{Vuik:2018}%
  \BibitemOpen
  \bibfield  {author} {\bibinfo {author} {\bibfnamefont {A.}~\bibnamefont
  {{Vuik}}}, \bibinfo {author} {\bibfnamefont {B.}~\bibnamefont {{Nijholt}}},
  \bibinfo {author} {\bibfnamefont {A.~R.}\ \bibnamefont {{Akhmerov}}}, \ and\
  \bibinfo {author} {\bibfnamefont {M.}~\bibnamefont {{Wimmer}}},\ }\href@noop
  {} {\ }\Eprint {http://arxiv.org/abs/1806.02801} {arXiv:1806.02801}
  \BibitemShut {NoStop}%
\bibitem [{\citenamefont {{Avila}}\ \emph {et~al.}()\citenamefont {{Avila}},
  \citenamefont {{Pe{\~n}aranda}}, \citenamefont {{Prada}}, \citenamefont
  {{San-Jose}},\ and\ \citenamefont {{Aguado}}}]{Avila:2018}%
  \BibitemOpen
  \bibfield  {author} {\bibinfo {author} {\bibfnamefont {J.}~\bibnamefont
  {{Avila}}}, \bibinfo {author} {\bibfnamefont {F.}~\bibnamefont
  {{Pe{\~n}aranda}}}, \bibinfo {author} {\bibfnamefont {E.}~\bibnamefont
  {{Prada}}}, \bibinfo {author} {\bibfnamefont {P.}~\bibnamefont {{San-Jose}}},
  \ and\ \bibinfo {author} {\bibfnamefont {R.}~\bibnamefont {{Aguado}}},\
  }\href {https://arxiv.org/abs/1807.04677} {\ }\Eprint
  {http://arxiv.org/abs/1807.04677} {arXiv:1807.04677} \BibitemShut {NoStop}%
\bibitem [{\citenamefont {Hansen}\ \emph {et~al.}(2018)\citenamefont {Hansen},
  \citenamefont {Danon},\ and\ \citenamefont {Flensberg}}]{Hansen:2018}%
  \BibitemOpen
  \bibfield  {author} {\bibinfo {author} {\bibfnamefont {E.~B.}\ \bibnamefont
  {Hansen}}, \bibinfo {author} {\bibfnamefont {J.}~\bibnamefont {Danon}}, \
  and\ \bibinfo {author} {\bibfnamefont {K.}~\bibnamefont {Flensberg}},\ }\href
  {https://link.aps.org/doi/10.1103/PhysRevB.97.041411} {\bibfield  {journal}
  {\bibinfo  {journal} {Phys. Rev. B}\ }\textbf {\bibinfo {volume} {{\bf
  97}}},\ \bibinfo {pages} {041411} (\bibinfo {year} {2018})}\BibitemShut
  {NoStop}%
\bibitem [{\citenamefont {Pe{\~n}aranda}\ \emph {et~al.}()\citenamefont
  {Pe{\~n}aranda}, \citenamefont {Aguado}, \citenamefont {San-Jose},\ and\
  \citenamefont {Prada}}]{Elsa2}%
  \BibitemOpen
  \bibfield  {author} {\bibinfo {author} {\bibfnamefont {F.}~\bibnamefont
  {Pe{\~n}aranda}}, \bibinfo {author} {\bibfnamefont {R.}~\bibnamefont
  {Aguado}}, \bibinfo {author} {\bibfnamefont {P.}~\bibnamefont {San-Jose}}, \
  and\ \bibinfo {author} {\bibfnamefont {E.}~\bibnamefont {Prada}},\
  }\href@noop {} {\ }\Eprint {http://arxiv.org/abs/1807.11924}
  {arXiv:1807.11924} \BibitemShut {NoStop}%
\bibitem [{\citenamefont {Kells}\ \emph {et~al.}(2012)\citenamefont {Kells},
  \citenamefont {Meidan},\ and\ \citenamefont {Brouwer}}]{Kells:2012}%
  \BibitemOpen
  \bibfield  {author} {\bibinfo {author} {\bibfnamefont {G.}~\bibnamefont
  {Kells}}, \bibinfo {author} {\bibfnamefont {D.}~\bibnamefont {Meidan}}, \
  and\ \bibinfo {author} {\bibfnamefont {P.~W.}\ \bibnamefont {Brouwer}},\
  }\href {https://link.aps.org/doi/10.1103/PhysRevB.86.100503} {\bibfield
  {journal} {\bibinfo  {journal} {Phys. Rev. B}\ }\textbf {\bibinfo {volume}
  {{\bf 86}}},\ \bibinfo {pages} {100503} (\bibinfo {year} {2012})}\BibitemShut
  {NoStop}%
\bibitem [{\citenamefont {Fleckenstein}\ \emph {et~al.}(2018)\citenamefont
  {Fleckenstein}, \citenamefont {Dom\'{\i}nguez}, \citenamefont
  {Traverso~Ziani},\ and\ \citenamefont {Trauzettel}}]{Fleckenstein:2018}%
  \BibitemOpen
  \bibfield  {author} {\bibinfo {author} {\bibfnamefont {C.}~\bibnamefont
  {Fleckenstein}}, \bibinfo {author} {\bibfnamefont {F.}~\bibnamefont
  {Dom\'{\i}nguez}}, \bibinfo {author} {\bibfnamefont {N.}~\bibnamefont
  {Traverso~Ziani}}, \ and\ \bibinfo {author} {\bibfnamefont {B.}~\bibnamefont
  {Trauzettel}},\ }\href {https://link.aps.org/doi/10.1103/PhysRevB.97.155425}
  {\bibfield  {journal} {\bibinfo  {journal} {Phys. Rev. B}\ }\textbf {\bibinfo
  {volume} {{\bf 97}}},\ \bibinfo {pages} {155425} (\bibinfo {year}
  {2018})}\BibitemShut {NoStop}%
\bibitem [{\citenamefont {Aseev}\ \emph {et~al.}(2018)\citenamefont {Aseev},
  \citenamefont {Klinovaja},\ and\ \citenamefont {Loss}}]{Aseev:2018}%
  \BibitemOpen
  \bibfield  {author} {\bibinfo {author} {\bibfnamefont {P.~P.}\ \bibnamefont
  {Aseev}}, \bibinfo {author} {\bibfnamefont {J.}~\bibnamefont {Klinovaja}}, \
  and\ \bibinfo {author} {\bibfnamefont {D.}~\bibnamefont {Loss}},\ }\href
  {https://link.aps.org/doi/10.1103/PhysRevB.98.155414} {\bibfield  {journal}
  {\bibinfo  {journal} {Phys. Rev. B}\ }\textbf {\bibinfo {volume} {{\bf
  98}}},\ \bibinfo {pages} {155414} (\bibinfo {year} {2018})}\BibitemShut
  {NoStop}%
\bibitem [{\citenamefont {Chevallier}\ \emph {et~al.}(2012)\citenamefont
  {Chevallier}, \citenamefont {Sticlet}, \citenamefont {Simon},\ and\
  \citenamefont {Bena}}]{Pascal1}%
  \BibitemOpen
  \bibfield  {author} {\bibinfo {author} {\bibfnamefont {D.}~\bibnamefont
  {Chevallier}}, \bibinfo {author} {\bibfnamefont {D.}~\bibnamefont {Sticlet}},
  \bibinfo {author} {\bibfnamefont {P.}~\bibnamefont {Simon}}, \ and\ \bibinfo
  {author} {\bibfnamefont {C.}~\bibnamefont {Bena}},\ }\href
  {https://link.aps.org/doi/10.1103/PhysRevB.85.235307} {\bibfield  {journal}
  {\bibinfo  {journal} {Phys. Rev. B}\ }\textbf {\bibinfo {volume} {{\bf
  85}}},\ \bibinfo {pages} {235307} (\bibinfo {year} {2012})}\BibitemShut
  {NoStop}%
\bibitem [{\citenamefont {Chevallier}\ \emph {et~al.}(2013)\citenamefont
  {Chevallier}, \citenamefont {Simon},\ and\ \citenamefont {Bena}}]{Pascal2}%
  \BibitemOpen
  \bibfield  {author} {\bibinfo {author} {\bibfnamefont {D.}~\bibnamefont
  {Chevallier}}, \bibinfo {author} {\bibfnamefont {P.}~\bibnamefont {Simon}}, \
  and\ \bibinfo {author} {\bibfnamefont {C.}~\bibnamefont {Bena}},\ }\href
  {https://link.aps.org/doi/10.1103/PhysRevB.88.165401} {\bibfield  {journal}
  {\bibinfo  {journal} {Phys. Rev. B}\ }\textbf {\bibinfo {volume} {{\bf
  88}}},\ \bibinfo {pages} {165401} (\bibinfo {year} {2013})}\BibitemShut
  {NoStop}%
\bibitem [{\citenamefont {Cayao}\ \emph {et~al.}(2015)\citenamefont {Cayao},
  \citenamefont {Prada}, \citenamefont {San-Jose},\ and\ \citenamefont
  {Aguado}}]{Cayao:2015}%
  \BibitemOpen
  \bibfield  {author} {\bibinfo {author} {\bibfnamefont {J.}~\bibnamefont
  {Cayao}}, \bibinfo {author} {\bibfnamefont {E.}~\bibnamefont {Prada}},
  \bibinfo {author} {\bibfnamefont {P.}~\bibnamefont {San-Jose}}, \ and\
  \bibinfo {author} {\bibfnamefont {R.}~\bibnamefont {Aguado}},\ }\href
  {https://link.aps.org/doi/10.1103/PhysRevB.91.024514} {\bibfield  {journal}
  {\bibinfo  {journal} {Phys. Rev. B}\ }\textbf {\bibinfo {volume} {{\bf
  91}}},\ \bibinfo {pages} {024514} (\bibinfo {year} {2015})}\BibitemShut
  {NoStop}%
\bibitem [{\citenamefont {Huang}\ \emph {et~al.}(2018)\citenamefont {Huang},
  \citenamefont {Pan}, \citenamefont {Liu}, \citenamefont {Sau}, \citenamefont
  {Stanescu},\ and\ \citenamefont {Das~Sarma}}]{Huang:2018}%
  \BibitemOpen
  \bibfield  {author} {\bibinfo {author} {\bibfnamefont {Y.}~\bibnamefont
  {Huang}}, \bibinfo {author} {\bibfnamefont {H.}~\bibnamefont {Pan}}, \bibinfo
  {author} {\bibfnamefont {C.-X.}\ \bibnamefont {Liu}}, \bibinfo {author}
  {\bibfnamefont {J.~D.}\ \bibnamefont {Sau}}, \bibinfo {author} {\bibfnamefont
  {T.~D.}\ \bibnamefont {Stanescu}}, \ and\ \bibinfo {author} {\bibfnamefont
  {S.}~\bibnamefont {Das~Sarma}},\ }\href
  {https://link.aps.org/doi/10.1103/PhysRevB.98.144511} {\bibfield  {journal}
  {\bibinfo  {journal} {Phys. Rev. B}\ }\textbf {\bibinfo {volume} {{\bf
  98}}},\ \bibinfo {pages} {144511} (\bibinfo {year} {2018})}\BibitemShut
  {NoStop}%
\bibitem [{\citenamefont {Reeg}\ and\ \citenamefont
  {Maslov}(2015)}]{Reeg:2015}%
  \BibitemOpen
  \bibfield  {author} {\bibinfo {author} {\bibfnamefont {C.~R.}\ \bibnamefont
  {Reeg}}\ and\ \bibinfo {author} {\bibfnamefont {D.~L.}\ \bibnamefont
  {Maslov}},\ }\href {https://link.aps.org/doi/10.1103/PhysRevB.92.134512}
  {\bibfield  {journal} {\bibinfo  {journal} {Phys. Rev. B}\ }\textbf {\bibinfo
  {volume} {{\bf 92}}},\ \bibinfo {pages} {134512} (\bibinfo {year}
  {2015})}\BibitemShut {NoStop}%
\bibitem [{\citenamefont {Klinovaja}\ and\ \citenamefont
  {Loss}(2015)}]{Klinovaja:2015}%
  \BibitemOpen
  \bibfield  {author} {\bibinfo {author} {\bibfnamefont {J.}~\bibnamefont
  {Klinovaja}}\ and\ \bibinfo {author} {\bibfnamefont {D.}~\bibnamefont
  {Loss}},\ }\href {http://dx.doi.org/10.1140/epjb/e2015-50882-2} {\bibfield
  {journal} {\bibinfo  {journal} {Eur. Phys. J. B}\ }\textbf {\bibinfo {volume}
  {{\bf 88}}},\ \bibinfo {pages} {62} (\bibinfo {year} {2015})}\BibitemShut
  {NoStop}%
\bibitem [{ste()}]{step}%
  \BibitemOpen
  \href@noop {} {}\bibinfo {note} {We take all parameters to follow a
  step-function profile due to the atomically sharp boundary of the
  superconductor that proximitizes the nanowire.}\BibitemShut {Stop}%
\bibitem [{neg()}]{neglect}%
  \BibitemOpen
  \href@noop {} {}\bibinfo {note} {In reality, the spin-orbit interaction and
  $g$-factor are nonzero in the proximitized section, but these quantities are
  expected to be much smaller than their corresponding values in the quantum
  dot. Therefore, we neglect them not only for analytical simplicity but also
  to emphasize the fact that a pinned ABS can appear in a model in which
  topological superconductivity is not possible.}\BibitemShut {Stop}%
\bibitem [{\citenamefont {Dmytruk}\ and\ \citenamefont
  {Klinovaja}(2018)}]{olesia}%
  \BibitemOpen
  \bibfield  {author} {\bibinfo {author} {\bibfnamefont {O.}~\bibnamefont
  {Dmytruk}}\ and\ \bibinfo {author} {\bibfnamefont {J.}~\bibnamefont
  {Klinovaja}},\ }\href {https://link.aps.org/doi/10.1103/PhysRevB.97.155409}
  {\bibfield  {journal} {\bibinfo  {journal} {Phys. Rev. B}\ }\textbf {\bibinfo
  {volume} {{\bf 97}}},\ \bibinfo {pages} {155409} (\bibinfo {year}
  {2018})}\BibitemShut {NoStop}%
\bibitem [{\citenamefont {Lim}\ \emph {et~al.}(2013)\citenamefont {Lim},
  \citenamefont {Lopez},\ and\ \citenamefont {Serra}}]{serra}%
  \BibitemOpen
  \bibfield  {author} {\bibinfo {author} {\bibfnamefont {J.~S.}\ \bibnamefont
  {Lim}}, \bibinfo {author} {\bibfnamefont {R.}~\bibnamefont {Lopez}}, \ and\
  \bibinfo {author} {\bibfnamefont {L.}~\bibnamefont {Serra}},\ }\href
  {http://stacks.iop.org/0295-5075/103/i=3/a=37004} {\bibfield  {journal}
  {\bibinfo  {journal} {Europhys. Lett.}\ }\textbf {\bibinfo {volume} {{\bf
  103}}},\ \bibinfo {pages} {37004} (\bibinfo {year} {2013})}\BibitemShut
  {NoStop}%
\bibitem [{\citenamefont {Nijholt}\ and\ \citenamefont
  {Akhmerov}(2016)}]{anton}%
  \BibitemOpen
  \bibfield  {author} {\bibinfo {author} {\bibfnamefont {B.}~\bibnamefont
  {Nijholt}}\ and\ \bibinfo {author} {\bibfnamefont {A.~R.}\ \bibnamefont
  {Akhmerov}},\ }\href {https://link.aps.org/doi/10.1103/PhysRevB.93.235434}
  {\bibfield  {journal} {\bibinfo  {journal} {Phys. Rev. B}\ }\textbf {\bibinfo
  {volume} {{\bf 93}}},\ \bibinfo {pages} {235434} (\bibinfo {year}
  {2016})}\BibitemShut {NoStop}%
\bibitem [{\citenamefont {Klinovaja}\ and\ \citenamefont
  {Loss}(2012)}]{compose}%
  \BibitemOpen
  \bibfield  {author} {\bibinfo {author} {\bibfnamefont {J.}~\bibnamefont
  {Klinovaja}}\ and\ \bibinfo {author} {\bibfnamefont {D.}~\bibnamefont
  {Loss}},\ }\href {http://link.aps.org/doi/10.1103/PhysRevB.86.085408}
  {\bibfield  {journal} {\bibinfo  {journal} {Phys. Rev. B}\ }\textbf {\bibinfo
  {volume} {{\bf 86}}},\ \bibinfo {pages} {085408} (\bibinfo {year}
  {2012})}\BibitemShut {NoStop}%
\bibitem [{\citenamefont {Klinovaja}\ \emph {et~al.}(2012)\citenamefont
  {Klinovaja}, \citenamefont {Stano},\ and\ \citenamefont {Loss}}]{Peter}%
  \BibitemOpen
  \bibfield  {author} {\bibinfo {author} {\bibfnamefont {J.}~\bibnamefont
  {Klinovaja}}, \bibinfo {author} {\bibfnamefont {P.}~\bibnamefont {Stano}}, \
  and\ \bibinfo {author} {\bibfnamefont {D.}~\bibnamefont {Loss}},\ }\href
  {http://link.aps.org/doi/10.1103/PhysRevLett.109.236801} {\bibfield
  {journal} {\bibinfo  {journal} {Phys. Rev. Lett.}\ }\textbf {\bibinfo
  {volume} {{\bf 109}}},\ \bibinfo {pages} {236801} (\bibinfo {year}
  {2012})}\BibitemShut {NoStop}%
\bibitem [{\citenamefont {de~Gennes}\ and\ \citenamefont
  {Saint-James}(1963)}]{deGennes:1963}%
  \BibitemOpen
  \bibfield  {author} {\bibinfo {author} {\bibfnamefont {P.~G.}\ \bibnamefont
  {de~Gennes}}\ and\ \bibinfo {author} {\bibfnamefont {D.}~\bibnamefont
  {Saint-James}},\ }\href {\doibase
  http://dx.doi.org/10.1016/0031-9163(63)90148-3} {\bibfield  {journal}
  {\bibinfo  {journal} {Phys. Lett.}\ }\textbf {\bibinfo {volume} {{\bf 4}}},\
  \bibinfo {pages} {151} (\bibinfo {year} {1963})}\BibitemShut {NoStop}%
\bibitem [{\citenamefont {Dmytruk}\ \emph {et~al.}(2018)\citenamefont
  {Dmytruk}, \citenamefont {Chevallier}, \citenamefont {Loss},\ and\
  \citenamefont {Klinovaja}}]{olesia2}%
  \BibitemOpen
  \bibfield  {author} {\bibinfo {author} {\bibfnamefont {O.}~\bibnamefont
  {Dmytruk}}, \bibinfo {author} {\bibfnamefont {D.}~\bibnamefont {Chevallier}},
  \bibinfo {author} {\bibfnamefont {D.}~\bibnamefont {Loss}}, \ and\ \bibinfo
  {author} {\bibfnamefont {J.}~\bibnamefont {Klinovaja}},\ }\href
  {https://link.aps.org/doi/10.1103/PhysRevB.98.165403} {\bibfield  {journal}
  {\bibinfo  {journal} {Phys. Rev. B}\ }\textbf {\bibinfo {volume} {{\bf
  98}}},\ \bibinfo {pages} {165403} (\bibinfo {year} {2018})}\BibitemShut
  {NoStop}%
\bibitem [{ine()}]{inequality}%
  \BibitemOpen
  \href@noop {} {}\bibinfo {note} {As the field enters through
  $\exp(\Delta_ZL/\alpha)$, a strong inequality $\Delta_Z\gg\alpha/L$ is not
  necessarily needed. In practice, $\Delta_Z\gtrsim\alpha/L$ should suffice, an
  important distinction when considering the range of field strengths that are
  experimentally accessible.}\BibitemShut {Stop}%
\bibitem [{nom()}]{nomismatch}%
  \BibitemOpen
  \href@noop {} {}\bibinfo {note} {Note that in the absence of Fermi velocity
  mismatch ($v_F=\alpha$), where there are no Fabry-Perot oscillations to
  lowest order in $1/\alpha$, we in fact see that no resonance condition is
  required to have a pinned ABS. In this case, the ABS energy is given by
  $E=2\Delta e^{-\Delta_ZL/\alpha}$ for arbitrary $L$ in the limit
  $\Delta_ZL/\alpha\gg1$. However, for an exact solution, it is not possible to
  completely eliminate Fabry-Perot oscillations, and we have confirmed
  numerically that a resonance condition must still be satisfied to observe a
  pinned ABS.}\BibitemShut {Stop}%
\bibitem [{spi()}]{spinx}%
  \BibitemOpen
  \href@noop {} {}\bibinfo {note} {It can be shown numerically that the ABS
  remains pinned near zero energy because its spin polarization in the
  direction of the magnetic field becomes negligible. Thus, the energy of ABS
  depends only weakly on the applied magnetic field.}\BibitemShut {Stop}%
\bibitem [{dis()}]{disorder}%
  \BibitemOpen
  \href@noop {} {}\bibinfo {note} {Note that disorder within the
  superconducting layer does not affect the ABS provided that the mean free
  path is $\ell\sim d$, as shown in Ref.~\cite{Reeg:2018}. As the low-energy
  pinning of the ABS requires tuning the chemical potential to the middle of
  the Zeeman gap, disorder within the nanowire does not affect the stability of
  this pinning behavior if on-site fluctuations in the chemical potential are
  smaller than $\Delta_Z$. This situation should be distinguished from one in
  which strong disorder pins ABSs to zero energy
  \cite{Patrick,Diego}.}\BibitemShut {Stop}%
\bibitem [{num()}]{numerics}%
  \BibitemOpen
  \href@noop {} {}\bibinfo {note} {While not shown, we also checked numerically
  that the pinning of the ABS is insensitive to variations in $L_s$, $\mu_s$,
  and $\Delta_0$, provided that we maintain the assumptions
  $L_s\gg2\sqrt{\mu_st_s}/\Delta_0$ and $\mu_s\gg\Delta_0$.}\BibitemShut
  {Stop}%
\bibitem [{\citenamefont {Liu}\ \emph {et~al.}(2012)\citenamefont {Liu},
  \citenamefont {Potter}, \citenamefont {Law},\ and\ \citenamefont
  {Lee}}]{Patrick}%
  \BibitemOpen
  \bibfield  {author} {\bibinfo {author} {\bibfnamefont {J.}~\bibnamefont
  {Liu}}, \bibinfo {author} {\bibfnamefont {A.~C.}\ \bibnamefont {Potter}},
  \bibinfo {author} {\bibfnamefont {K.~T.}\ \bibnamefont {Law}}, \ and\
  \bibinfo {author} {\bibfnamefont {P.~A.}\ \bibnamefont {Lee}},\ }\href
  {https://link.aps.org/doi/10.1103/PhysRevLett.109.267002} {\bibfield
  {journal} {\bibinfo  {journal} {Phys. Rev. Lett.}\ }\textbf {\bibinfo
  {volume} {{\bf 109}}},\ \bibinfo {pages} {267002} (\bibinfo {year}
  {2012})}\BibitemShut {NoStop}%
\bibitem [{\citenamefont {Rainis}\ \emph {et~al.}(2013)\citenamefont {Rainis},
  \citenamefont {Trifunovic}, \citenamefont {Klinovaja},\ and\ \citenamefont
  {Loss}}]{Diego}%
  \BibitemOpen
  \bibfield  {author} {\bibinfo {author} {\bibfnamefont {D.}~\bibnamefont
  {Rainis}}, \bibinfo {author} {\bibfnamefont {L.}~\bibnamefont {Trifunovic}},
  \bibinfo {author} {\bibfnamefont {J.}~\bibnamefont {Klinovaja}}, \ and\
  \bibinfo {author} {\bibfnamefont {D.}~\bibnamefont {Loss}},\ }\href
  {http://link.aps.org/doi/10.1103/PhysRevB.87.024515} {\bibfield  {journal}
  {\bibinfo  {journal} {Phys. Rev. B}\ }\textbf {\bibinfo {volume} {{\bf
  87}}},\ \bibinfo {pages} {024515} (\bibinfo {year} {2013})}\BibitemShut
  {NoStop}%
\bibitem [{\citenamefont {Law}\ \emph {et~al.}(2009)\citenamefont {Law},
  \citenamefont {Lee},\ and\ \citenamefont {Ng}}]{law}%
  \BibitemOpen
  \bibfield  {author} {\bibinfo {author} {\bibfnamefont {K.~T.}\ \bibnamefont
  {Law}}, \bibinfo {author} {\bibfnamefont {P.~A.}\ \bibnamefont {Lee}}, \ and\
  \bibinfo {author} {\bibfnamefont {T.~K.}\ \bibnamefont {Ng}},\ }\href
  {https://link.aps.org/doi/10.1103/PhysRevLett.103.237001} {\bibfield
  {journal} {\bibinfo  {journal} {Phys. Rev. Lett.}\ }\textbf {\bibinfo
  {volume} {{\bf 103}}},\ \bibinfo {pages} {237001} (\bibinfo {year}
  {2009})}\BibitemShut {NoStop}%
\bibitem [{\citenamefont {Wimmer}\ \emph {et~al.}(2011)\citenamefont {Wimmer},
  \citenamefont {Akhmerov}, \citenamefont {Dahlhaus},\ and\ \citenamefont
  {Beenakker}}]{wimmer}%
  \BibitemOpen
  \bibfield  {author} {\bibinfo {author} {\bibfnamefont {M.}~\bibnamefont
  {Wimmer}}, \bibinfo {author} {\bibfnamefont {A.~R.}\ \bibnamefont
  {Akhmerov}}, \bibinfo {author} {\bibfnamefont {J.~P.}\ \bibnamefont
  {Dahlhaus}}, \ and\ \bibinfo {author} {\bibfnamefont {C.~W.~J.}\ \bibnamefont
  {Beenakker}},\ }\href {http://stacks.iop.org/1367-2630/13/i=5/a=053016}
  {\bibfield  {journal} {\bibinfo  {journal} {New J. Phys.}\ }\textbf {\bibinfo
  {volume} {{\bf 13}}},\ \bibinfo {pages} {053016} (\bibinfo {year}
  {2011})}\BibitemShut {NoStop}%
\bibitem [{\citenamefont {Chevallier}\ and\ \citenamefont
  {Klinovaja}(2016)}]{denis2016}%
  \BibitemOpen
  \bibfield  {author} {\bibinfo {author} {\bibfnamefont {D.}~\bibnamefont
  {Chevallier}}\ and\ \bibinfo {author} {\bibfnamefont {J.}~\bibnamefont
  {Klinovaja}},\ }\href {https://link.aps.org/doi/10.1103/PhysRevB.94.035417}
  {\bibfield  {journal} {\bibinfo  {journal} {Phys. Rev. B}\ }\textbf {\bibinfo
  {volume} {{\bf 94}}},\ \bibinfo {pages} {035417} (\bibinfo {year}
  {2016})}\BibitemShut {NoStop}%
\bibitem [{\citenamefont {{Grivnin}}\ \emph {et~al.}()\citenamefont
  {{Grivnin}}, \citenamefont {{Bor}}, \citenamefont {{Heiblum}}, \citenamefont
  {{Oreg}},\ and\ \citenamefont {{Shtrikman}}}]{Grivnin:2018}%
  \BibitemOpen
  \bibfield  {author} {\bibinfo {author} {\bibfnamefont {A.}~\bibnamefont
  {{Grivnin}}}, \bibinfo {author} {\bibfnamefont {E.}~\bibnamefont {{Bor}}},
  \bibinfo {author} {\bibfnamefont {M.}~\bibnamefont {{Heiblum}}}, \bibinfo
  {author} {\bibfnamefont {Y.}~\bibnamefont {{Oreg}}}, \ and\ \bibinfo {author}
  {\bibfnamefont {H.}~\bibnamefont {{Shtrikman}}},\ }\href
  {https://arxiv.org/abs/1807.06632} {\ }\Eprint
  {http://arxiv.org/abs/1807.06632} {arXiv:1807.06632} \BibitemShut {NoStop}%
\bibitem [{bul()}]{bulk}%
  \BibitemOpen
  \href@noop {} {}\bibinfo {note} {However, due to the potential ambiguity of a
  purely local probe of MBSs, it would also be beneficial to supplement any
  local measurements with additional probes of any nonlocal or bulk signatures
  of a topological phase transition, e.g., as suggested in
  Refs.~\cite{Hansen:2018,Akhmerov:2011,Fregoso:2013,Szumniak:2017,Rosdahl:2018,Serina:2018,
  Schrade:2018}.}\BibitemShut {Stop}%
\bibitem [{\citenamefont {Prada}\ \emph {et~al.}(2012)\citenamefont {Prada},
  \citenamefont {San-Jose},\ and\ \citenamefont {Aguado}}]{madrid}%
  \BibitemOpen
  \bibfield  {author} {\bibinfo {author} {\bibfnamefont {E.}~\bibnamefont
  {Prada}}, \bibinfo {author} {\bibfnamefont {P.}~\bibnamefont {San-Jose}}, \
  and\ \bibinfo {author} {\bibfnamefont {R.}~\bibnamefont {Aguado}},\ }\href
  {https://link.aps.org/doi/10.1103/PhysRevB.86.180503} {\bibfield  {journal}
  {\bibinfo  {journal} {Phys. Rev. B}\ }\textbf {\bibinfo {volume} {{\bf
  86}}},\ \bibinfo {pages} {180503} (\bibinfo {year} {2012})}\BibitemShut
  {NoStop}%
\bibitem [{\citenamefont {Akhmerov}\ \emph {et~al.}(2011)\citenamefont
  {Akhmerov}, \citenamefont {Dahlhaus}, \citenamefont {Hassler}, \citenamefont
  {Wimmer},\ and\ \citenamefont {Beenakker}}]{Akhmerov:2011}%
  \BibitemOpen
  \bibfield  {author} {\bibinfo {author} {\bibfnamefont {A.~R.}\ \bibnamefont
  {Akhmerov}}, \bibinfo {author} {\bibfnamefont {J.~P.}\ \bibnamefont
  {Dahlhaus}}, \bibinfo {author} {\bibfnamefont {F.}~\bibnamefont {Hassler}},
  \bibinfo {author} {\bibfnamefont {M.}~\bibnamefont {Wimmer}}, \ and\ \bibinfo
  {author} {\bibfnamefont {C.~W.~J.}\ \bibnamefont {Beenakker}},\ }\href
  {https://link.aps.org/doi/10.1103/PhysRevLett.106.057001} {\bibfield
  {journal} {\bibinfo  {journal} {Phys. Rev. Lett.}\ }\textbf {\bibinfo
  {volume} {{\bf 106}}},\ \bibinfo {pages} {057001} (\bibinfo {year}
  {2011})}\BibitemShut {NoStop}%
\bibitem [{\citenamefont {Fregoso}\ \emph {et~al.}(2013)\citenamefont
  {Fregoso}, \citenamefont {Lobos},\ and\ \citenamefont
  {Das~Sarma}}]{Fregoso:2013}%
  \BibitemOpen
  \bibfield  {author} {\bibinfo {author} {\bibfnamefont {B.~M.}\ \bibnamefont
  {Fregoso}}, \bibinfo {author} {\bibfnamefont {A.~M.}\ \bibnamefont {Lobos}},
  \ and\ \bibinfo {author} {\bibfnamefont {S.}~\bibnamefont {Das~Sarma}},\
  }\href {https://link.aps.org/doi/10.1103/PhysRevB.88.180507} {\bibfield
  {journal} {\bibinfo  {journal} {Phys. Rev. B}\ }\textbf {\bibinfo {volume}
  {{\bf 88}}},\ \bibinfo {pages} {180507} (\bibinfo {year} {2013})}\BibitemShut
  {NoStop}%
\bibitem [{\citenamefont {Szumniak}\ \emph {et~al.}(2017)\citenamefont
  {Szumniak}, \citenamefont {Chevallier}, \citenamefont {Loss},\ and\
  \citenamefont {Klinovaja}}]{Szumniak:2017}%
  \BibitemOpen
  \bibfield  {author} {\bibinfo {author} {\bibfnamefont {P.}~\bibnamefont
  {Szumniak}}, \bibinfo {author} {\bibfnamefont {D.}~\bibnamefont
  {Chevallier}}, \bibinfo {author} {\bibfnamefont {D.}~\bibnamefont {Loss}}, \
  and\ \bibinfo {author} {\bibfnamefont {J.}~\bibnamefont {Klinovaja}},\ }\href
  {https://link.aps.org/doi/10.1103/PhysRevB.96.041401} {\bibfield  {journal}
  {\bibinfo  {journal} {Phys. Rev. B}\ }\textbf {\bibinfo {volume} {{\bf
  96}}},\ \bibinfo {pages} {041401} (\bibinfo {year} {2017})}\BibitemShut
  {NoStop}%
\bibitem [{\citenamefont {Rosdahl}\ \emph {et~al.}(2018)\citenamefont
  {Rosdahl}, \citenamefont {Vuik}, \citenamefont {Kjaergaard},\ and\
  \citenamefont {Akhmerov}}]{Rosdahl:2018}%
  \BibitemOpen
  \bibfield  {author} {\bibinfo {author} {\bibfnamefont {T.~O.}\ \bibnamefont
  {Rosdahl}}, \bibinfo {author} {\bibfnamefont {A.}~\bibnamefont {Vuik}},
  \bibinfo {author} {\bibfnamefont {M.}~\bibnamefont {Kjaergaard}}, \ and\
  \bibinfo {author} {\bibfnamefont {A.~R.}\ \bibnamefont {Akhmerov}},\ }\href
  {https://link.aps.org/doi/10.1103/PhysRevB.97.045421} {\bibfield  {journal}
  {\bibinfo  {journal} {Phys. Rev. B}\ }\textbf {\bibinfo {volume} {{\bf
  97}}},\ \bibinfo {pages} {045421} (\bibinfo {year} {2018})}\BibitemShut
  {NoStop}%
\bibitem [{\citenamefont {Serina}\ \emph {et~al.}(2018)\citenamefont {Serina},
  \citenamefont {Loss},\ and\ \citenamefont {Klinovaja}}]{Serina:2018}%
  \BibitemOpen
  \bibfield  {author} {\bibinfo {author} {\bibfnamefont {M.}~\bibnamefont
  {Serina}}, \bibinfo {author} {\bibfnamefont {D.}~\bibnamefont {Loss}}, \ and\
  \bibinfo {author} {\bibfnamefont {J.}~\bibnamefont {Klinovaja}},\ }\href
  {https://link.aps.org/doi/10.1103/PhysRevB.98.035419} {\bibfield  {journal}
  {\bibinfo  {journal} {Phys. Rev. B}\ }\textbf {\bibinfo {volume} {{\bf
  98}}},\ \bibinfo {pages} {035419} (\bibinfo {year} {2018})}\BibitemShut
  {NoStop}%
\bibitem [{\citenamefont {Schrade}\ and\ \citenamefont {Fu}()}]{Schrade:2018}%
  \BibitemOpen
  \bibfield  {author} {\bibinfo {author} {\bibfnamefont {C.}~\bibnamefont
  {Schrade}}\ and\ \bibinfo {author} {\bibfnamefont {L.}~\bibnamefont {Fu}},\
  }\href {https://arxiv.org/abs/1809.06370} {\ }\Eprint
  {http://arxiv.org/abs/1809.06370} {arXiv:1809.06370} \BibitemShut {NoStop}%
\end{thebibliography}%

\end{document}